\documentclass[fleqn,usenatbib]{mnras}

\usepackage{newtxtext,newtxmath}
\usepackage[T1]{fontenc}
\usepackage{ae,aecompl}

\usepackage{graphicx}	% Including figure files
\usepackage{amsmath}	% Advanced maths commands
\usepackage{amssymb}	% Extra maths symbols
\usepackage{widetext}

\usepackage{amsmath}
\usepackage{graphicx}
\usepackage[colorinlistoftodos]{todonotes}
\usepackage{bm}
\newcommand{\norm}[1]{\left\lVert#1\right\rVert}

\newcommand{\VEV}[1]{\left\langle #1 \right\rangle}

\newcommand{\veck}{\mathbf{k}}
\newcommand{\vecr}{\mathbf{r}}
\newcommand{\hatr}{\mathbf{\hat{r}}}
\newcommand{\pc}[1]{\textcolor{green}{[#1]}}
\DeclareMathOperator{\sinc}{sinc}

\title[Full-Sky Lensing Reconstruction]{Full-Sky Lensing Reconstruction of 21 cm Intensity Maps}

\author[Chakraborty \& Pullen]{
Priyesh Chakraborty,$^{1}$ Anthony R. Pullen$^{1}$\thanks{E-mail: anthony.pullen@nyu.edu}
\\
$^{1}$Center for Cosmology and Particle Physics, Department of Physics, New York University, 726 Broadway, New York, NY, 10003, U.S.A.\\
}

\date{Accepted XXX. Received YYY; in original form ZZZ}

\pubyear{2018}

\begin{document}
\label{firstpage}
\pagerange{\pageref{firstpage}--\pageref{lastpage}}
\maketitle

\begin{abstract}
Weak gravitational lensing of the 21 cm radiation is expected to be an important cosmological probe {for post-reionization} physics. We investigate the reconstruction of the matter density perturbations using a quadratic minimum variance estimator. The next generation of {line} intensity mapping ({LIM}) surveys such as HIRAX and CHIME will cover a larger sky fraction, which requires one to account for the curvature in the sky. Thus, we extend the plane-parallel flat-sky formalism for lensing reconstruction to account for a full-sky survey using the Spherical Fourier-Bessel (SFB) expansion. Using the HIRAX 21 cm survey as a basis, we make predictions for lensing-reconstruction noise in our formalism and compare our results with the predictions from the plane-parallel formalism. We find agreement with the plane-parallel noise power spectrum at small scales and a significant deviation at scales {$L\lesssim \ell_{\rm res}-k_{\rm eq}R$ where $R$ is the radius of the shell volume, $k_{\rm eq}$ is the wavenumber for matter-radiation equality, and $\ell_{\rm res}$ is the angular resolution scale}. Furthermore, we derive the SFB flat-sky reconstruction noise and compare it with the full-sky SFB case as well as the plane-parallel case, finding minor deviations from the full-sky noise due to sphericity.  We also determine that, {in the absence of non-Gaussian statistics of the intensity field but accounting for foregrounds,} the signal-to-noise ratio (SNR) for $C_\ell^{\phi\phi}$ using our SFB estimator increases by {107\%}.  This shows that accounting for the curved sky in {LIM} weak lensing will be crucial for large-scale cosmology.
\end{abstract}


%%%%%%%%%%%%%%%%%%%%%%%%%%%%%%%%%%%%%%%%%%%%%%%%%%

%%%%%%%%%%%%%%%%% BODY OF PAPER %%%%%%%%%%%%%%%%%%

\section{Introduction}

Weak gravitational lensing by large-scale structure (LSS) \citep{1987A&A...184....1B,1988A&A...206..199L,1989MNRAS.239..195C} (see \citet{2017SchpJ..1232440B} for a recent review article) is a powerful and ubiquitous probe of cosmology.  As an unbiased tracer of LSS it can be used to measure the distribution of matter \citep{2018MNRAS.475.3165C} as well as the growth of structure \citep{2015APh....63...23H}, an important probe of gravity.  The two common tracers of the gravitational lensing potential are shear deformations in the shapes of galaxies {\citep{2018PhRvD..98d2005P,2018arXiv180909148H}} and {deflections in the cosmic microwave background (CMB) that sample both the convergence and shear fields \citep{2018arXiv180706210P,2017ApJ...849..124O,PhysRevD.95.123529}.  A recent addition to the lensing tracers is the cosmic infrared background (CIB) \citep{2018PhRvD..97l3539S}, in which the authors derive the optimal flat-sky lensing estimator for the CIB while accounting for the CIB trispectrum and self-lensing by low-redshift perturbations.} The lensing potential also serves as a foreground for the $B$-mode CMB polarization through the lensing of the $E$-mode polarization field \citep{2016ARA&A..54..227K}.

Line intensity mapping (IM) \citep{1990MNRAS.247..510S,1997ApJ...475..429M,1999ApJ...512..547S,2008PhRvL.100i1303C,2008MNRAS.383.1195W} has emerged as a new tracer of large-scale structure, in which the aggregate line emission from star-forming galaxies and the intergalactic medium (IGM) is tabulated into three-dimensional maps and has both astrophysical and cosmological applications.  The most promising emission line being considered is the 21 cm line from neutral hydrogen, but several other lines are also being considered, such as CO, CII, and Ly$\alpha$ (see \citet{2017arXiv170909066K} for the latest review on these lines).  {LIM} has lately been considered as a promising new probe of the lensing potential.  {LIM} lensing has mostly been considered in terms of the 3D Fourier-space quadratic estimator derived in \citet{2006ApJ...653..922Z} while numerous other estimators have also been considered \citep{2006ApJ...647..719M,2007MNRAS.381..447M,2008MNRAS.388.1819L,2009MNRAS.394..704M,2010PhRvD..81l3015L,2014MNRAS.439L..36P,2015MNRAS.448.2368P,2018MNRAS.474.1787R}  LIM has significant advantages over traditional lensing tracers.  For one, unlike CMB fluctuations, 21-cm fluctuations do not experience Silk damping on small scales \citep{1997ApJ...479..568H}; they are limited instead by the Jeans length of the gas which occurs at much smaller scales \citep{2004PhRvL..92u1301L}, providing more power to sample the lens field. Also, the CMB has only one source plane, while intensity maps will have multiple source planes which can sample the same lenses, greatly increasing the potential to constrain the gravitational perturbation field.  Shear maps from galaxy lensing can also have multiple source planes, but these maps tend to use photometric redshifts which are more uncertain and limit the number of source planes that can be applied.  Intensity maps are also optimized to reach redshifts all the way out through the reionization epoch and beyond, greatly extending the reach of lensing surveys.  One disadvantage of intensity maps for lensing compared to CMB maps is that while the CMB is nearly Gaussian, intensity maps at lower redshifts will be non-Gaussian due to nonlinear clustering.  {However, recent work has determined various alternatives that can reduce this effect, including bias-hardened estimators \citep{2018JCAP...07..046F} and accounting for the trispectra \citep{2018PhRvD..97l3539S}.  Also, the foreground contamination for 21-cm intensity maps is orders of magnitude larger than for the CMB, which will degrade the ability to map the lensing potential.}

The formalism for weak lensing of intensity maps using a quadratic estimator {such as} in \citet{2006ApJ...653..922Z} is typically written in terms of the \emph{plane-parallel formalism}, where the 3D intensity maps are transformed into Fourier modes in terms of $\mathbf{k}=(k_\parallel,\mathbf{k}_\perp)$, where $k_\parallel$ is the component of the wavevector $\mathbf{k}$ along the line of sight, and $\mathbf{k}_\perp$ is the transverse component.  This formalism is ideal for surveys over small areas such that the line of sight does not change significantly over the survey.  However, some upcoming 21-cm surveys are currently being designed to survey very large areas.  For example, the HIRAX 21-cm survey \citep{2016SPIE.9906E..5XN} will probe redshifts $0.8<z<2.5$ and map all the Southern Sky.  In this limit, we cannot assume the line-of-sight is constant over the survey.  In fact, we expect that Fourier analysis of intensity maps in this limit is not optimal, and that a new basis is required that accounts for the curvature of the sky.

A basis that serves this function is the \emph{spherical Fourier-Bessel basis} (SFB).  In this basis, the 3D map is decomposed into modes labeled by $(k,\ell)$,  This basis has a rich history in cosmological theory and analysis, particularly for galaxy clustering and weak lensing \citep{1991MNRAS.249..678B,1994MNRAS.266..219F,1995MNRAS.272..885F,1994ApJ...423L..93L,1995ApJ...449..446Z,1995MNRAS.275..483H,2005PhRvD..72b3516C,2012arXiv1211.0310L,2012A&A...540A.115R,2012MNRAS.422.2341S,2012A&A...540A..60L,2013MNRAS.436.3792P,2013PhRvD..88b3502Y}.  
Recently, \citet{Liu:2016xzv} proposed analyzing intensity maps in the SFB basis, and \citet{2018MNRAS.476.4403C} showed that SFB moments of 3D cosmological maps can be used directly to estimate the matter power spectrum in Fourier space. This paper will describe a powerful application for the SFB basis, namely weak gravitational lensing.  We expect the SFB and plane-parallel {lensing} formalisms to deviate {due to distorted correlations of the intensity field in mostly angular modes ($\ell\sim kR$)}, scales at which lensing can help answer cosmological questions.  One such concern is constraining gravity, which has significant large-scale effects and for which lensing is an unbiased probe.  These being large-scale effects imply that a SFB formalism is vital to modeling the effects in data correctly.%  {In addition, while in the plane-parallel formalism lensing only correlates modes with different $k_\perp$ values, in the full sky picture lensing also correlates modes with different $k_\parallel$ values, which appear as correlated modes with different $k$ values in the SFB formalism.  This will increase the sampling of modes of the lensing potential which could increase the statistical power of the signal.}

In this paper, we construct the formalism for estimating the gravitational potential from 3D intensity maps over the full sky, {assuming Gaussian statistics in the intensity field.}   After reviewing the weak lensing estimator in the plane-parallel formalism as well as the spherical Fourier-Bessel (SFB) decomposition, we derive a minimum-variance quadratic estimator for the spherical harmonic moments of the weak lensing potential given the SFB series of a 3D intensity map.  With this estimator, we then derive an expression for the noise bias of the angular power spectrum of the lensing potential $N_\ell^{\phi\phi}$ in this formalism.  We also present the $N_\ell^{\phi\phi}$ prediction using the SFB flat-sky limit.  We then compare both SFB predictions for $N_\ell^{\phi\phi}$ to the prediction from \cite{2006ApJ...653..922Z} using the plane-parallel formalism, assuming instrumental properties of the HIRAX 21 cm survey as a test case.

We find that both SFB predictions agree with the plane-parallel prediction at small scales but deviate dramatically at scales {$L$ for the lensing potential where the estimator is sourced by some scales that are mostly angular in the intensity map with high signal-to-noise ratios.  This corresponds to $L\lesssim \ell_{\rm res}-k_{\rm eq}R$ where $R$ is the radius of the shell volume, $k_{\rm eq}$ is the peak of the matter power spectrum set by the wavenumber that enters the cosmic horizon at matter-radiation equality, and $\ell_{\rm res}$ is the angular scale where the instrument beam increases the noise in angular harmonic space by 25\%.  Note that the full-sky and flat-sky SFB predictions deviate from the plane-parallel prediction at the same scale, while the full-sky SFB prediction deviates slightly from the flat-sky prediction for $\ell\lesssim10$,} similar to the full-sky vs. flat-sky deviation for CMB lensing.  We then determine {the sensitivity} of $C_\ell^{\phi\phi}$ and $C_\ell^{\kappa g}$, the convergence-galaxy angular cross-power spectrum, for the HIRAX 21-cm intensity mapping survey \citep{2016SPIE.9906E..5XN} {over a redshift range $1.38\leq z\leq 2.57$} using our new formalism, assuming the galaxy sample that will be produced by the Large Synoptic Survey Telescope (LSST) \citep{2012arXiv1211.0310L}.  With a range of {\textbf{$100<L_{\rm dev}<200$}} for the scale where $N_\ell^{\phi\phi}$ deviates from the plane-parallel prediction, we find that the SNR for $C_\ell^{\phi\phi}$ using our SFB estimator increases by {107\%}, while $C_\ell^{\kappa g}$ with galaxies from the upcoming Large Synoptic Survey Telescope {only has a modest increase of 10\% with much higher deviations if considering just large angular scales.  Note that these results will change once non-Gaussian intensity fields due to nonlinear clustering are taken into account; we leave this for future work.}  These results suggest that at large angular scales intensity mapping {could} be a much more powerful probe of large-scale structure than previously thought, which could have many applications including {improving} measurements of gravity on the largest scales.

The plan of our paper is as follows: in Section \ref{S:Clpp} we review the form of the gravitational lensing potential and the noise bias from intensity map estimates in the plane-parallel formalism.  In Section \ref{S:sfbform} we review the SFB formalism including our formalism for power spectrum predictions on a spherical shell.  We present our full-sky $N_\ell^{\phi\phi}$ derivation in Section \ref{S:clppfull} and compare this result to the plane-parallel prediction in Section \ref{S:compare}.  We conclude in Section \ref{S:conc}.  We assume cosmological parameters consistent with the Planck 2015 cosmology \citep{2016A&A...594A..13P}, namely $\Omega_bh^2=0.02225$, $\Omega_mh^2=0.1419$, $h=0.676$, $\Omega_\nu h^2=0.00064$, $N_{\rm eff}=3.046$, $n_s=0.964$, and $A_s=2.2\times10^{-9}$.

\iffalse

\section{Preliminary Definitions}\label{S:prelim}
We have assumed a $\Lambda$CDM cosmology with comoving coordinates $\vec{\textbf{r}}=(\hat{\textbf{r}},r)$ where $\hat{\textbf{r}}$ represents the unit angular vector and the comoving distance $r=r(z)$ at redshift $z$ is given by:
\begin{eqnarray}
r(z)=\frac{c}{H_0}\int_{0}^{z(\nu)}\frac{dz'}{E(z')}
\end{eqnarray}
where $c$ is the speed of light and $H_0$ is the Hubble parameter today and:
\begin{eqnarray}
1+z=\frac{\nu_0}{\nu}\, ,\qquad E(z)=\sqrt{\Omega_{\Lambda}+\Omega_{m}(1+z)^3}
\end{eqnarray}
where $\nu_0$ is the rest frequency of the 21 cm spectral line and $\Omega_{\Lambda}$ and $\Omega_{m}$ are standard dark energy and matter density parameters.\newline
Let the 21 cm intensity field be $\theta(\hat{\textbf{r}},\nu)$ and the lensed field be $\widetilde{\theta}(\hat{\textbf{r}},r)$. For a lensing displacement field $\phi(\hat{\textbf{r}})$:
\begin{eqnarray}
\widetilde{\theta}(\hat{\textbf{r}},r)=\theta(\hat{\textbf{r}}+\nabla\phi,r)\text{  ,  } \phi(\mathbf{\hat{r}})=\frac{2}{c^2}\int_0^{\chi_s}d\chi\frac{\chi_s-\chi}{\chi_s\chi}\Phi(\chi_s\mathbf{\hat{r}},z)
\end{eqnarray}
where $\psi$ is the scalar metric perturbation field and $W(r)$ is the relevant lensing kernel. Above and henceforth, tildes will refer to lensed quantities.

\fi

\section{Angular Power Spectrum of Lensing Potential} \label{S:Clpp}

Let the 21 cm intensity field be {$T(\hat{\textbf{r}},r)$} and the lensed field be {$\widetilde{T}(\hat{\textbf{r}},r)$}. For a lensing potential $\phi(\hat{\textbf{r}})$,
{\begin{eqnarray}
\widetilde{T}(\hat{\textbf{r}},r)=T(\hat{\textbf{r}}+\nabla\phi,r).
\end{eqnarray}
We will use the conventions where the Fourier transform is given by
\begin{eqnarray}
T(\veck)&=&\int d^3x\,e^{-i\veck\cdot\vecr}T(\vecr)\nonumber\\
T(\vecr)&=&\int\frac{d^3k}{(2\pi)^3}e^{i\veck\cdot\vecr}T(\veck)\, ,
\end{eqnarray}
the covariance between Fourier modes is written as
\begin{eqnarray}
\VEV{T(\veck)T^*(\veck')}=(2\pi)^3\delta_D^3(\veck-\veck')P_T(k)
\end{eqnarray}
where $P_T(k)$ is the power spectrum of the brightness temperature fluctuations and $k=|\mathbf{k}|$ is the mode wavenumber.}

The lensing potential, which describes the weak lensing of extragalactic photons due to gravity, has been given extensively in the literature in the form
\begin{eqnarray}\label{E:phi}
\phi(\mathbf{\hat{r}})=\frac{2}{c^2}\int_0^{\chi_s}d\chi\frac{\chi_s-\chi}{\chi_s\chi}\Phi(\chi_s\mathbf{\hat{r}},z)\, ,
\end{eqnarray}
where $\chi$ and $\chi_s$ are the comoving distances of the lenses and source plane.  $\Phi$ is the gravitational potential, given in the Poisson equation as
{\begin{eqnarray}
\nabla^2\Phi=\frac{3}{2}\Omega_{m,0}H_0^2(1+z)\delta\, ,
\end{eqnarray}}
with $\Omega_{m,0}$ being the matter density today and $\delta$ being the matter overdensity.  Inserting this into Eq.~\ref{E:phi} and expanding $\delta$ in Fourier components, we can write the angular power spectrum for $\phi$ as
\begin{eqnarray}
C_\ell^{\phi\phi}=\frac{9}{c^4}\Omega_{m,0}^2H_0^4\frac{2}{\pi}\int\frac{dk}{k^2}[W_\ell^\phi(k)]^2P(k)\, ,
\end{eqnarray}
where {$P(k)$ is the matter power spectrum,}
\begin{eqnarray}
W_\ell^\phi(k)=\int_0^{\chi_s}d\chi\,\frac{\chi_s-\chi}{\chi_s\chi}(1+z)D(z)j_\ell(k\chi)\, ,
\end{eqnarray}
and $D(z)$ is the growth function.%  Since the lensing kernel is broad, we can use the Limber approximation, which gives us
%\begin{eqnarray}
%C_\ell^{\phi\phi}=\frac{9}{c^4}\Omega_{m,0}^2H_0^4\int_0^{\chi_s}d\chi\frac{\chi^2}{\ell^4}\left(\frac{\chi_s-\chi}{\chi_s\chi}\right)^2(1+z)^2P\left(\frac{\ell}{\chi},z\right)
%\end{eqnarray}

The estimator for the lensing potential in an intensity mapping context was first derived in \citet{2006ApJ...653..922Z} within a plane-parallel formalism.  In order to account for a finite volume, $k_\parallel$ {is} discretized as $k_{\parallel,j}=2\pi j/\Delta R$ with $j=1,2,...$ and $\Delta R$ being the depth of the survey.  In addition, $\mathbf{k}_\perp$ was rewritten as $\mathbf{k}_\perp=\boldsymbol{\ell}/R$ with $R$ being the distance to the source.  In this formalism, the noise of the measured auto-power spectrum of the lensing potential $\phi$ at angular scale $\ell$ is given by
\begin{eqnarray}\label{E:nlpp}
N_L^{\phi\phi}=\left[\sum_{j\geq j_{\rm min}}\int \frac{d^2\ell}{(2\pi)^2}\frac{f_\phi^2(\mathbf{L},\boldsymbol{\ell},k_{\parallel,j})}{C_\ell^{\rm tot}(k_{\parallel,j})C_{|\mathbf{L}-\boldsymbol{\ell}|}^{\rm tot}(k_{\parallel,j})}\right]^{-1}\, ,
\end{eqnarray}
where
\begin{eqnarray}
f_\phi(\mathbf{L},\boldsymbol{\ell},k_\parallel)&=&\mathbf{L}\cdot\boldsymbol{\ell}C_\ell(k_\parallel)+\mathbf{L}\cdot(\mathbf{L}-\boldsymbol{\ell})C_{|\mathbf{L}-\boldsymbol{\ell}|}(k_\parallel)\nonumber\\
C_\ell(k_\parallel)&=&{\frac{[1+\beta(k_{\parallel}/k)^2]^2}{R^2\Delta R}} P_T\left(k=\sqrt{\left(\frac{\ell}{R}\right)^2+k_\parallel^2}\right)\nonumber\\
C_\ell^{\rm tot}(k_\parallel)&=&C_\ell(k_\parallel)+C_\ell^{\rm N}\, ,
\end{eqnarray}
{ where $\beta=f/b$ is the growth rate divided by the linear bias and $j_{\rm min}$ denotes the minimum $k_{\parallel,j}$ that is not contaminated by foregrounds. Note that $C_{\ell}^{\rm N}$ is the noise power spectrum which will be defined in Sec.~\ref{S:compare}}

\section{Spherical Fourier Bessel Formalism} \label{S:sfbform}

In this section we review the formalism for the SFB basis including a continuous $k$ and discrete $k$ values, after which {we} will derive the covariance of the discrete moments in terms of the power spectrum.  
\iffalse
We will use the conventions where the Fourier transform is given by
\begin{eqnarray}
T(\veck)&=&\int d^3x\,e^{-i\veck\cdot\vecr}T(\vecr)\nonumber\\
T(\vecr)&=&\int\frac{d^3k}{(2\pi)^3}e^{i\veck\cdot\vecr}T(\veck)\, ,
\end{eqnarray}
the covariance between Fourier modes is written as
\begin{eqnarray}
\VEV{T(\veck)T^*(\veck')}=(2\pi)^3\delta_D^3(\veck-\veck')P(k)
\end{eqnarray}
where $P(k)$ is the power spectrum of the brightness temperature fluctuations and $k=|\mathbf{k}|$ is the mode wavenumber.
\fi

In the LIM context, a cosmology is assumed in order to relate the radial distance $r$ to redshift $z$ in the form 
\begin{eqnarray}
r(z)=\int_0^z\frac{dz'}{H(z')}\, ,
\end{eqnarray}
where $H(z)$ is the Hubble rate in terms of the relative densities of matter and dark energy.  The redshift $z$ is then related to the observed frequency of the survey line, such that $\nu=\nu_{\rm rest}/(1+z)$.  In the linearized form, $z$ can be directly related to $r$ over a narrow ranged centered at $r_{\rm ref}(z_{\rm ref})$ according to
\begin{eqnarray}\label{E:rnu}
r=r_{\rm ref}-\frac{c(1+z_{\rm ref})^2}{\nu_{\rm rest}H(z_{\rm ref})}(\nu-\nu_{\rm ref})\, .
\end{eqnarray}

\subsection{Spherical Fourier Bessel Transform Review}

In \cite{Liu:2016xzv}, the SFB transform in three dimensions is given in the form
\begin{eqnarray}
T_{\ell m}(k)=\sqrt{\frac{2}{\pi}}\int d^3r\,j_\ell(kr)Y_{\ell m}^*(\mathbf{\hat{r}})T(\mathbf{r})\, ,
\end{eqnarray}
where $k=|\mathbf{k}|$ is the mode wavenumber, $r=|\mathbf{r}|$ is the distance along the line-of-sight, $\mathbf{\hat{r}}=\mathbf{r}/r$ is the unit direction vector, $\ell$ and $m$ are the degree and order, respectively, of the spherical harmonic function $Y_{\ell m}(\mathbf{\hat{r}})$, and $j_\ell(x)$ is the spherical Bessel function of the first kind.  The inverse of this transform is
\begin{eqnarray}\label{E:tr}
T(\vecr)=\sqrt{\frac{2}{\pi}}\sum_{\ell m}\int dk\,k^2j_\ell(kr)Y_{\ell m}(\mathbf{\hat{r}})T_{\ell m}(k)\, .
\end{eqnarray}
Any function defined on $\mathcal{R}^3$ can be written in this basis.  A useful relation in this context is the identity
\begin{eqnarray}
\int dr\,r^2j_\ell(kr)j_\ell(k'r)=\frac{\pi}{2}\frac{\delta_D(k-k')}{k^2}\, .
\end{eqnarray}

It is shown in \cite{Liu:2016xzv} that for the case with an isotropic power spectrum {in the temperature fluctuations $P_T(k)=b^2\bar{T}^2P(k)$ where $b$ is the clustering bias and $\bar{T}$ is the mean brightness temperature}, the covariance between $T_{\ell m}(k)$ modes can be written as
\begin{eqnarray}
\VEV{T_{\ell m}(k)T_{\ell' m'}^*(k')}=P_T(k)\frac{\delta_D(k-k')}{k^2}\delta_{\ell\ell'}\delta_{mm'}\, ,
\end{eqnarray}
such that all the modes are independent. L16 also asserted that if rotational invariance of the fluctuations $T(\vecr)$ is broken, \emph{e.g.} redshift-space distortions, then the covariance between modes changes to
\begin{eqnarray}\label{E:tcl}
\VEV{T_{\ell m}(k)T_{\ell' m'}^*(k')}=C_\ell(k,k')\delta_{\ell\ell'}\delta_{mm'}\, .
\end{eqnarray}
In recent work \citep{2018MNRAS.476.4403C}, the SFB power spectrum including redshift-space distortions (RSD) has been shown to be given in the form
\begin{eqnarray}\label{E:clkk}
C_\ell(k_1,k_2)&=&\left(\frac{2}{\pi}\right)^2\int dk\,k^2P_T(k)[W^0_\ell(k_1,k)-\beta W^r_\ell(k_1,k)]\nonumber\\
&&\times[W^0_\ell(k_2,k)-\beta W^r_\ell(k_2,k)]\, ,
\end{eqnarray}
where the RSD parameter $\beta=f/b$ is the growth rate divided by the clustering bias and $W_\ell^0$ and $W_\ell^r$ are window functions given by
{
\begin{eqnarray}\label{E:rsd}
W^0_\ell(k,k')&=&\int dr\,r^2j_{\ell}(kr)j_\ell(k'r)\nonumber\\
&=& \frac{\pi}{2}\frac{\delta_D(k-k')}{k^2}\, , \nonumber \\
W^r_\ell(k,k')&=&\int dr\,r^2j_\ell(kr)j_\ell''(k'r)\, ,
\end{eqnarray}
}
where the second derivative of $j_\ell(x)$ can be written as
\begin{eqnarray}
j_\ell''(x)&=&\frac{\ell(\ell-1)}{(2\ell-1)(2\ell+1)}j_{\ell-2}(x)-\frac{2\ell^2+2\ell-1}{(2\ell-1)(2\ell+3)}j_\ell(x)\nonumber\\
&&+\frac{(\ell+1)(\ell+2)}{(2\ell+1)(2\ell+3)}j_{\ell+2}(x)\, .
\end{eqnarray}
Note that strictly speaking, $r$ is in redshift-space, not real-space, {but this should have a small effect once the small survey window along the line of sight, which is also in redshift space, is taken into account; thus we will not mention this further.}

For a partial volume all-sky survey defined by the window function $\phi(r)=1$ inside the survey and 0 outside, $W_\ell^0$ and $W_\ell^r$ in Eq.~\ref{E:rsd} are modulated by {$\phi(r)$} in the integrand, in which case $W_\ell^0$ will no longer be a delta function.
For an all-sky survey, L16 showed this leads to an estimator for $P(k)$ defined as
\begin{eqnarray}
S_\ell(k)\equiv2\pi^2\left[\int d^3r\phi^2(r)j_\ell^2(kr)\right]^{-1}\frac{\sum_m|T_{\ell m}^{\rm meas}(k)|^2}{2\ell+1}\, ,
\end{eqnarray}
where the $\ell$ modes can be combined to construct a minimum-variance estimator for $P(k)$.

\subsection{Spherical Fourier Bessel Series}

{The spherical Fourier Bessel (SFB) transform can be discretized into a spherical Fourier Bessel series in a finite volume.  This has typically been performed for a spherical volume \citep{1995MNRAS.272..885F,2012A&A...540A..60L,2013MNRAS.436.3792P}; however, more realistic surveys are constructed over spherical shells.  Thus we will define our survey volume as a spherical shell with inner radius $r_{\rm min}$ and outer radius $r_{\rm max}$ when modeling a transformation using the spherical Fourier Bessel series.}

{For the sphere, the wavenumber $k$ in the transform must have discrete values to satisfy a boundary condition at $T(r=r_{\rm max})$.  For a shell, a boundary condition at $T(r=r_{\rm min})$ must also be satisfied.  In our paper, we use Dirichlet boundary conditions $T=0$ at both boundaries.  Modeling the radial behavior as $j_\ell(kr)$ as in the SFB transform makes it impossible to satisfy both boundary conditions.  However, we instead choose $k$ values that satisfy $j_\ell(kr_{\rm max})=0$ exactly, then picking values of $k$ that most closely satisfy $j_\ell(kr_{\rm min})=0$.  This discrete SFB formalism is inspired by a full treatment for the shell which will be presented in \citet{samushia} and to which we had early access, in which $j_\ell(kr)$ is replaced by a linear combination of the spherical Bessel functions of the first and second kind, $j_\ell(kr)+Ay_\ell(kr)$, such that the boundary conditions at both $r_{\rm min}$ and $r_{\rm max}$ can be satisfied exactly.}

{We define moments for the map as $T_{\ell m n}$ given by
\begin{eqnarray}\label{E:tlmn}
T_{\ell m n}=\int_{r_{\rm min}\leq r\leq r_{\rm max}} d^3rj_\ell(k_{\ell n}r)Y_{\ell m}^*(\hatr)T(\vecr)\, ,
\end{eqnarray}
where $k_{\ell n}=s_{\ell n}/r_{\rm max}$ and $s_{\ell n}$ is the $n^{\rm th}$ root of both $j_\ell(x)$ and $j_\ell(ax)$ where $a=r_{\rm min}/r_{\rm max}$, such that $j_\ell(s_{\ell n})=0\,\forall n$, while approximately satisfying $j_\ell(as_{\ell n})=0$.  Note that for $a=0$ we attain the spherical case in which $j_\ell(as_{\ell n})=0$ is automatically satisfied for $\ell>0$, leaving only the boundary condition $j_\ell(s_{\ell n})=0\,\forall n$.  In this case we define $q_{\ell n}$ as the $n^{\rm th}$ root of only $j_\ell(x)$.}

{Here we derive the lowest $k_{\ell n}$ allowed in our basis as well as the spacing between consecutive $k_{\ell n}$s.  We use the approximation for the roots of the $j_\ell(x)$ given by $q_{\ell n}\simeq\ell+n\pi$ for $n=1,2,3,...$ \citep{2012A&A...540A..60L}.  We will also only consider the case in which the volume is a thin shell, where $\Delta R=r_{\rm max}-r_{\rm min}\ll r_{\rm max}$ or equivalently $1-a\ll 1$.   First we consider the case where $\ell\ll n\pi$ to find the lowest $s_{\ell 1}=q_{\ell n}$ by setting $n$.  In this case we can approximate $q_{\ell n}\simeq n\pi$.  Since the roots of $j_\ell(x)$ are approximately $\pi$ apart, and we need both $q_{\ell n}$ and $aq_{\ell n}$ to be roots, we can pick these roots to have the shortest possible difference of $\pi$.  This leads to the condition that $n=1/(1-a)=r_{\rm max}/\Delta R$.  Thus the first value for $s_{\ell n}$ is
$s_{\ell1}=\ell+\pi r_{\rm max}/\Delta R$.  To get subsequent $s_{\ell n}$s, given an $s_{\ell n}$ we can always define $q_{\ell n'}'=s_{\ell n}+n'\pi$ as the set of subsequent roots of $j_\ell(x)$.  We can then repeat the previous exercise showing that the next root of both $j_\ell(x)$ and $j_\ell(ax)$ is a distance $\pi r_{\rm max}/\Delta R$ away.  Thus in the case $\ell\ll n\pi$ we can set
\begin{eqnarray}
s_{\ell n}\simeq\ell+n\pi\frac{r_{\rm max}}{\Delta R}\, ,
\end{eqnarray}
where $n=1,2,3,...$ and the spacing between $k_{\ell n}$s is $\Delta k=\pi/\Delta R$.}

{Next we consider the case $\ell\gtrsim n\pi$.  Now the first roots for $j_\ell(x)$ and $j_\ell(ax)$ are approximately $\ell$ and $a\ell$, respectively.  We then want to find a common root.  Since $a<1$ we should find the subsequent roots for $j_\ell(ax)$; since $\ell\gg\pi$, one of these roots will probably be equal to the first root of $j_\ell(x)$, namely $x\simeq\ell$ where $\ell\gg 1$.  Thus we set $\ell=a(\ell+n\pi)$, giving us $n=\ell(1-a)/(\pi a)$.  Substituting this in $q_{\ell n}$ we find $s_{\ell1}=\ell/a$.  Getting subsequent $s_{\ell n}$s for this case is not trivial.  Unlike the case where $\ell\ll n\pi$, we cannot set all subsequent roots of both $j_\ell(x)$ and $j_\ell(ax)$ to be multiples of $\pi$ away from $s_{\ell1}$ since they are inherently offset.  In this regime the $k$-spacing can be much smaller than $\pi/\Delta R$.  Also in this regime the $x$ values that satisfy both $j_\ell(x)$ and $j_\ell(ax)$ tend to be well in between the $q_{\ell n}$s, causing the chosen $s_{\ell n}$s not to satisfy $j_\ell(as_{\ell n})$ very well.  This is unavoidable as long as our basis functions only consist of $j_\ell$s.  However, at high enough $n$ the offset mentioned earlier will become negligible, and from then on the $k_{\ell n}$s will be spaced as $\Delta k=\pi/\Delta R$.}

{Note that in both cases since $s_{\ell1}=k_{\ell1}r_{\rm max}>\ell$, the SFB basis does not probe modes $kr_{\rm max}<\ell$.  This is similar to L16 in which modes $kr_{\rm max}<\ell$ have extremely low signal-to-noise for power spectrum measurements and correspond to modes $k_\perp>k$ in the plane-parallel limit, which is unphysical. In addition, the boundary condition alone does not account for only using an intensity map within the shell; we fully account for this in our theoretical derivations by only integrating Eq.~\ref{E:tlmn} in the region $r_{\rm min}\leq r\leq r_{\rm max}$.}

{The orthogonal conditions for the shell's SFB series are 
\begin{eqnarray}\label{E:jj1}
\int_{r_{\rm min}}^{r_{\rm max}}dr\,r^2j_\ell(k_{\ell n}r)j_{\ell}(k_{\ell n'}r)\simeq\frac{\delta_{nn'}}{\tau_{\ell n}}\, ,
\end{eqnarray}
and
\begin{eqnarray}\label{E:jj2}
\sum_n\tau_{\ell n}j_\ell(k_{\ell n}r)j_{\ell}(k_{\ell n}r')=\frac{\delta(r-r')}{r^2}\, ,
\end{eqnarray}
where
\begin{eqnarray}
\tau_{\ell n}^{-1}&=&\frac{r_{\rm max}^3}{2}[j_{\ell+1}(k_{\ell n}r_{\rm max})]^2-\frac{r_{\rm min}^3}{2}[j_{\ell}(k_{\ell n}r_{\rm min})]^2\nonumber\\
&&+\frac{r_{\rm min}^3}{2}j_{\ell-1}(k_{\ell n}r_{\rm min})j_{\ell+1}(k_{\ell n}r_{\rm min})\, .
\end{eqnarray}
Note that since $j_{\ell}(k_{\ell n}r_{\rm min})$ sometimes differs from zero a fair amount for $\ell\gg \pi n$, the integral in Eq.~\ref{E:jj1} and the sum in Eq.~\ref{E:jj2} are only approximately diagonal.  This allows the series for $T(\vecr)$ to be expanded as
\begin{eqnarray}
T(\vecr)=\sum_{\ell m n}\tau_{\ell n}j_\ell(k_{\ell n}r)Y_{\ell m}(\hatr)T_{\ell m n}\, .
\end{eqnarray}}

The SFB transform basis and the SFB series basis both contain all the information on $T(\mathbf{r})$ within the radial range {$r_{\rm min}<r<r_{\rm max}$}.  The discrete $k_{\ell n}$'s Nyquist sample the map in such a way that continuous $k$'s do not contain more (or less) information with a finite correlation length $\Delta k=\pi/\Delta R$.

We can write the moments for the SFB series $T_{\ell m n}$ in terms of the moments for the SFB transform $T_{\ell m}(k)$ by inserting Eq.~\ref{E:tr} in Eq.~\ref{E:tlmn} to find
\begin{eqnarray}\label{E:tlmntlmk}
T_{\ell m n}=\sqrt{\frac{2}{\pi}}\int dk\,k^2W_{\ell n}(k)T_{\ell m}(k)\, ,
\end{eqnarray}
where $W_{\ell n}(k)$ is given by
{\begin{eqnarray}\label{E:wlm}
W_{\ell n}(k)&\equiv&\int_{r_{\rm min}}^{r_{\rm max}} dr\,r^2j_\ell(k_{\ell n}r)j_\ell(kr)\nonumber\\
&=&\frac{r_{\rm max}^2k_{\ell n}j_{\ell-1}(k_{\ell n}r_{\rm max})j_\ell(kr_{\rm max})}{k^2-k_{\ell n}^2}\nonumber\\
&&-\frac{r_{\rm min}^2k_{\ell n}j_{\ell-1}(k_{\ell n}r_{\rm min})j_\ell(kr_{\rm min})}{k^2-k_{\ell n}^2}\nonumber\\
&&+\frac{r_{\rm min}^2kj_{\ell-1}(kr_{\rm min})j_\ell(k_{\ell n}r_{\rm min})}{k^2-k_{\ell n}^2}\, .
\end{eqnarray}}
For $k_{\ell n}r_{\rm max}\gg\ell$, $W_{\ell n}(k)$ is a highly-peaked function at $k=k_{\ell n}$, {and for all cases} $W_{\ell n}(k_{\ell n})=1/\tau_{\ell n}$.

With these expressions, the covariance among modes $T_{\ell m n}$ can be written, assuming an intrinsic, isotropic power spectrum, as
\begin{eqnarray}\label{E:tlmtlm}
\VEV{T_{\ell m n}T_{\ell'm'n'}^*}=C_{\ell nn'}\delta_{\ell\ell'}\delta_{mm'}\, ,
\end{eqnarray}
where
\begin{eqnarray}
C_{\ell nn'}=\frac{2}{\pi}\int dk\,k^2\int dk'\,k'^2W_{\ell n}(k)W_{\ell n'}(k')C_\ell(k,k')\, .
\end{eqnarray}
{Note that the even in the isotropic $P(k)$ case the covariance matrix $C_{\ell nn'}$ is only approximately diagonal in $n$.  This is just due to the survey volume being a finite region.  This effect also occurs for the SFB transform in a finite volume.}  Inserting Eq.~\ref{E:clkk}, we can write the above equation in the form
\begin{eqnarray}\label{E:clnn}
C_{\ell nn'}&=&\frac{2}{\pi}\int dk\,k^2P_T(k)[W^0_{\ell n}(k)-\beta W^r_{\ell n}(k)]\nonumber\\
&&\times[W^0_{\ell n'}(k)-\beta W^r_{\ell n'}(k)]\, ,
\end{eqnarray}
where
{\begin{eqnarray}
W^0_{\ell n}(k)&=&W_{\ell n}(k)\nonumber\\
W^r_{\ell n}(k)&=&\int_{r_{\rm min}}^{r_{\rm max}} dr\,r^2j_\ell(k_{\ell n}r)j_\ell''(kr)\, ,
\end{eqnarray}}
Adding instrumental noise and effects from the instrumental beam and spectral window functions, this expression augments to
\begin{eqnarray}
\VEV{T_{\ell m n}^{\rm obs}T_{\ell' m' n'}^{\rm obs,*}}=\left[W_{\ell nn'}^{AB}C_{\ell nn'}+N_{\ell n n'}\right]\delta_{\ell\ell'}\delta_{mm'}\, ,
\end{eqnarray}
where we define $W_{\ell nn'}^{AB}\equiv W_{\ell nn'}^{A}W_\ell^{B}$ such that
\begin{eqnarray}\label{E:wna}
W_{\ell nn'}^A=A(k_{\ell n})A(k_{\ell n'})\, ,
\end{eqnarray}
where
\begin{eqnarray}\label{E:acalc}
A(k)=\frac{\pi}{4}\left[{\rm sinc}\left(\frac{k\Delta r-\pi}{2}\right)+{\rm sinc}\left(\frac{k\Delta r+\pi}{2}\right)\right]\, ,
\end{eqnarray}
and $\Delta r$ is the comoving distance corresponding to the spectral resolution and $W_\ell^B=e^{\ell^2\sigma_b^2}$, where $\sigma_b$ is the angular resolution.  This instrumental noise is given by
{\begin{eqnarray}
N_{\ell nn'}=\int_{r_{\rm min}}^{r_{\rm max}} dr\,r^2\Delta(r)j_\ell(k_{\ell n}r)j_\ell(k_{\ell n'}r)\sigma_N^2(r)\, ,
\end{eqnarray}}
where $z(r)$ is the redshift given comoving distance $r$, $\sigma_N(r)$ is the instrumental noise given as a function of $r$ which is related to frequency by Eq.~\ref{E:rnu}, and $\Delta(r)$ is the voxel size given by
\begin{eqnarray}
\Delta(r)=\Delta\Omega\left(\frac{\Delta\nu}{\nu}\right)r^2\frac{c[1+z(r)]}{ H[z(r)]}\, ,
\end{eqnarray}
where $\Delta\Omega$ and $\nu$ are the solid angle of the angular pixels and $\Delta \nu$ is the size of the frequency channels.  These instrumental effects are derived in Appendix \ref{S:intinst}.  {Note that since $\sigma(r)$ and $\Delta(r)$ are smooth functions over the shell, we can take them out of the integral setting them with the values $\sigma(R)$ and $\Delta(R)$ where $R=(r_{\rm min}+r_{\rm max})/2$.  With this approximation the integral can be performed analytically, allowing us to set
\begin{eqnarray}\label{E:Nln}
N_{\ell nn'}=\Delta(R)\sigma_N^2(R)\frac{\delta_{nn'}}{\tau_{\ell n}}\, .
\end{eqnarray}}

\subsection{SFB Flat-Sky Approximation}
{Here we consider a flat-sky approximation to the SFB basis in order to facilitate comparisons with the plane-parallel lensing formalism.  The flat-sky limit is characterized by $\theta\rightarrow 0$, i.e., small angles around the pole in angular coordinates on the sphere $(\theta,\varphi)$, which translates into $\ell\gg1$ in harmonic space. In the flat-sky coordinates, angular quantities are described by $\boldsymbol{\ell}=[\ell\cos(\varphi_\ell),\ell\sin(\varphi_\ell)]$, where $\ell$ is the continuum approximation of the Spherical Harmonic index. In this approximation, sums transition into integrals, for which we use $\int d^2\ell=\int_0^\infty \ell d\ell\int_0^{2\pi}d\varphi_\ell\approx\sum_\ell\ell\int_0^{2\pi}d\varphi_\ell$.%\newline
Inspired by the formalism from \citet{2005PhRvD..72b3516C} but converting to an SFB series formalism, we define the SFB flat-sky series transform as
\begin{eqnarray}
T_n(\boldsymbol{\ell})=\int_{r_{\rm min}\leq r\leq r_{\rm max}} dr\,r^2\int d^2\theta\,j_\ell(k_{\ell n}r)e^{-i\boldsymbol{\ell}\cdot\boldsymbol{\theta}}T(\mathbf{r})\, ,
\end{eqnarray}
and
\begin{eqnarray}
T(\vecr)=\sum_n\int\frac{d^2\ell}{(2\pi)^2}\tau_{\ell n}j_\ell(k_{\ell n}r)e^{i\boldsymbol{\ell}\cdot\boldsymbol{\theta}}T_n(\boldsymbol{\ell})\, .
\end{eqnarray}
We note the conversion from SFB full-sky moments to the flat-sky moments (and vice versa) are
\begin{eqnarray}
T_{\ell m n}&=&\sqrt[]{\frac{\ell}{2\pi}}i^{m}\int_0^{2\pi}\frac{d\varphi_{\ell}}{2\pi}e^{-im\varphi_{\ell}}T_n(\boldsymbol{\ell}) \nonumber \\
T_n(\boldsymbol{\ell})&=&\sqrt[]{\frac{2\pi}{\ell}}\sum_m i^{-m}e^{im\varphi_\ell}T_{\ell m n}\, .
\end{eqnarray}
Further, one can derive the covariance
\begin{eqnarray}
\langle T_n(\boldsymbol{\ell}) T_{n'}^*(\boldsymbol{\ell}')\rangle=(2\pi)^2C_{\ell nn'}\delta_D^2(\boldsymbol{\ell}-\boldsymbol{\ell}')\, .
\end{eqnarray}}

{It is evident that the SFB flat-sky basis is not strictly equivalent to the plane-parallel basis, given by
\begin{eqnarray}
T^{\rm obs}(k_{||,j},\boldsymbol{\ell})=R^2\int dx_{||}\,e^{-ik_{||,j}x_{||}}\int d^2\theta\,e^{-i\boldsymbol{\ell}\cdot\boldsymbol{\theta}}T(\mathbf{r})\, ,
\end{eqnarray}
with a power spectrum $\VEV{T^{\rm obs}(k_{||,j},\boldsymbol{\ell})T^{*,\rm obs}(k_{||,j'},\boldsymbol{\ell}')}=(2\pi)^2\delta^2(\boldsymbol{\ell}-\boldsymbol{\ell}')\delta_{jj'}(R^2\Delta R)^2\tilde{C}_\ell(k_{||,j})$ given by
\begin{eqnarray}
\tilde{C}_\ell(k_{||,j})=\frac{1}{(R\Delta R)^2}\int\frac{dk_{||}'}{2\pi}P_T(k_{||}',\boldsymbol{\ell})|W(k_{||}',k_{||,j})|^2\, ,
\end{eqnarray}
where
\begin{eqnarray}
W(k_{||}',k_{||,j})&=&\int_{-\Delta R/2}^{\Delta R/2}dx_{||}e^{i(k_{||}'-k_{||,j})x_{||}}\nonumber\\
&=&\Delta R\sinc[(k_{||}'-k_{||,j})\Delta R/2]\, .
\end{eqnarray}
However, it's useful to consider when the approximations
\begin{eqnarray}\label{E:sfb2pp}
\tau_{\ell n}C_{\ell n}&\simeq&R^2\Delta R\tilde{C}_\ell(k_{||,j})\simeq R^2\Delta R C_\ell(k_{||,j})\, ,
\end{eqnarray}
where $C_\ell(k_{||})=P_T(k_{||},\ell)/(R^2\Delta R)$ are true.  The approximation $\tilde{C}_\ell(k_{||,j})\simeq C_\ell(k_{||,j})$ is justified in that $W(k_{||}',k_{||,j})$ is highly peaked at $k_{||}'=k_{||,j}$ such that
\begin{eqnarray}
\tilde{C}_\ell(k_{||,j})&\simeq&\frac{P_T(k_{||,j},\ell)}{(R\Delta R)^2}\int\frac{dk_{||}'}{2\pi}|W(k_{||}',k_{||,j})|^2\nonumber\\
&=&\frac{P_T(k_{||,j},\ell)}{R^2\Delta R}=C_\ell(k_{||,j})\, .
\end{eqnarray}
This has allowed previous treatments of the weak lensing of intensity maps to set the power spectrum as $C_\ell(k_{||,j})$ and not $\tilde{C}_\ell(k_{||,j})$.}

{In the SFB case the approximation $\tau_{\ell n}C_{\ell n}\simeq R^2\Delta R C_\ell(k_{||,j})$ is only satisfied if the integrand of $C_{\ell n}$ is highly peaked, such that
\begin{eqnarray}
C_{\ell n}&\simeq&P_T(k_{\ell n},\boldsymbol{\ell})\frac{2}{\pi}\int dk\,k^2|W_{\ell n}(k)|^2\nonumber\\
&=&\frac{P_T(k_{\ell n},\boldsymbol{\ell})}{\tau_{\ell n}}=\frac{R^2\Delta R C_\ell(k_{||,j})}{\tau_{\ell n}}\, .
\end{eqnarray}
The condition that the integrand be highly peaked is satisfied when $\ell\ll k_{\ell n}R$ and $\ell\ll kR$ such that the asymptotic expansion of $j_\ell(x)$ can be applied and $W_{\ell n}(k)$ approaches a sinc function peaked at $k=k_{\ell n}$ of the form
\begin{eqnarray}
W_{\ell n}(k)\simeq\frac{\Delta R}{2k(k+k_{\ell n})}\sinc[(k-k_{\ell n})\Delta R/2]\cos[(k-k_{\ell n})R]
\end{eqnarray}
The condition $\ell\ll k_{\ell n}R$ is generally true for most $k_{\ell n}$, and $W_{\ell n}(k)$ tends to not get any other peaks even for $kR\lesssim\ell$. Thus we can remove $P_T(k)$ from the integral which leaves us with $\tau_{\ell n}C_{\ell n}\simeq P_T(k_{\ell n})$ for $k_{\ell n}R\gg\ell$.  However, for $k_{\ell n}R\sim\ell$, $W_{\ell n}(k)$ is not so highly peaked, causing mode mixing and giving us $\tau_{\ell n}C_{\ell n}<P_T(k_{\ell n})$.}

{Further, we wish to examine how $C_{\ell n}$ behaves at $\ell>k_{\rm eq}R$. Particularly, it is necessary to highlight the contribution of the integral from the region $kR<\ell$. In this regime, $j_{\ell}(kr)\approx \frac{(kr)^\ell}{(2\ell+1)!!}\sim\left(\frac{kr}{2\ell}\right)^\ell$, and thus,
\begin{eqnarray}
W_{\ell n}(k)&\sim&\int_{r_{\rm min}}^{r_{\rm max}} dr\,r^2\left(\frac{kr}{2\ell}\right)^\ell \frac{\sin{(k_{\ell n}r-\ell\pi/2)}}{k_{\ell n}r}\nonumber \\
&\sim&\frac{1}{k_{\ell n}}\left(\frac{k}{2\ell}\right)^\ell\int_{r_{\rm min}}^{r_{\rm max}}dr\,r^{\ell+1}\sin(k_{\ell n}r-\ell\pi/2)\nonumber \\
&\sim&\left(\frac{kR}{2\ell}\right)^\ell\int_{a}^{1}dx\,x^{\ell+1}\sin(q_{\ell n}x-\ell\pi/2)\nonumber \\
&\sim&\left(\frac{kR}{2\ell}\right)^\ell,
\end{eqnarray}
since the radial integrand here is bounded by $1$ and only contributes to suppression, the exact nature of which is irrelevant here. Thus we have shown that we may safely ignore the contribution from $kR<\ell$. Following the arguments for $k_{\ell n}R\gg \ell$, we have,
\begin{eqnarray}
\tau_{\ell n}C_{\ell n}&\approx&P_T(k_{\ell n}).
\end{eqnarray}
Indeed, this may be corroborated from Fig.~\ref{F:cln} for the specifications of the survey used in this paper (for which $P_T(k)$ is the matter-power spectrum). Particularly for the case $k_{\ell n}R>\ell>k_{\rm eq}R$, we have $P_T(k)\sim 1/k^2$, and given that $k_{\ell n}\sim \ell$, we may expect that $\tau_{\ell n}C_{\ell n}\sim \frac{1}{\ell^2}$, which is indeed suppressed.
}

\section{Full Sky weak lensing reconstruction}\label{S:clppfull}

{Here we derive the lensing estimator for a 21 cm intensity map, applying a formalism similar to that derived in \citet{2003PhRvD..67h3002O} for full-sky CMB lensing.}  We start with a 21 cm signal, perturbed weakly by a lensing potential.
Expanding the lensing equation to first order produces
{\begin{eqnarray}
\widetilde{T}(\hat{\textbf{r}},r) &=&T(\hat{\textbf{r}}+\nabla\phi,r) \\
 &\approx&T(\vec{\textbf{r}})+\nabla_j\phi\nabla^jT(\vec{\textbf{r}}) \, .
\end{eqnarray}}
Taking the SFB transform of this equation for the perturbation in the 21 cm field, {$\delta\widetilde{T}$}, produces its SFB moments
\begin{multline}\label{E:theta}
\delta\widetilde{T}_{\ell mn}=\sum_{\ell_1m_1}\sum_{\ell_2 m_2 n_2}\tau_{\ell_2 n_2}(-1)^{m}\left( {\begin{array}{ccc}
   \ell & \ell_1 & \ell_2\\
   -m & m_1 & m_2\\ 
   \end{array}} \right) \\
   \times W_{\ell\ell_2}^{nn_2}F_{\ell\ell_1\ell_2}T_{\ell_2 m_2n_2}\phi_{\ell_1 m_1}\, ,
\end{multline}
where $W$ is defined as

{\begin{eqnarray}\label{E:wllnn}
W_{\ell\ell'}^{nn'}&=&\int_{r_{\rm min}}^{r_{\rm max}} dr r^2 j_{\ell}(k_{\ell n}r)j_{\ell'}(k_{\ell'n'}r) \nonumber \\
&=&r_{\rm max}^3\int_a^1 dz z^2 j_{\ell}(q_{\ell n}z)j_{\ell'}(q_{\ell'n'}z)\, ,
\end{eqnarray}
where $a\equiv r_{\rm min}/r_{\rm max}$} and

\begin{multline}
F_{ll_1l_2}=[l_1(l_1+1)+l_2(l_2+1)-l(l+1)]\\
\times\sqrt[]{\frac{(2l_1+1)(2l+1)(2l_2+1)}{16\pi}}\left( {\begin{array}{ccc}
l & l_1 & l_2 \\
0 & 0 & 0 \\
\end{array}}\right)\, .
\end{multline}
In order to obtain this, we have used the well known identity:
\begin{multline}
\int d\hat{\textbf{r}}Y_{\ell m}^{*}(\hat{\textbf{r}})\nabla_{i}Y_{\ell_1 m_1}(\hat{\textbf{r}})\nabla^{i}Y_{\ell_2 m_2}(\hat{\textbf{r}})=(-1)^{m_1}\left( {\begin{array}{ccc}
l & l_1 & l_2 \\
-m & m_1 & m_2 \\
\end{array}}\right) \\ \times F_{\ell\ell_1\ell_2}\, .
\label{eq:3ylm}
\end{multline}

Note that $W_{\ell\ell'}^{nn'}$ serves as a coupling matrix that determines which {unlensed modes} source the lensed modes. {In order to determine the behavior of this coupling, we use the asymptotic approximation for $j_\ell(kr)$ at large $\ell$ in Appendix~\ref{A:saddle} given by
\begin{eqnarray}
j_{\ell}(kr)&\simeq&\frac{\sin{(k_{||}r+\varphi_{k}\ell-\pi\ell/2)}}{kr}\, ,
\end{eqnarray}
where $k_{||}=\sqrt{k_{\ell n}^2-(\ell/R)^2}$ and $\varphi_k={\rm arctan}(\ell/Rk_{||})$.  Now, upon integrating $W_{\ell\ell'}^{nn'}$ using this approximation, and after some reorganization and letting the (un)primed $k_{||}$ correspond to the (un)primed $n$ and $\ell$, one gets
\begin{widetext}
\begin{eqnarray}
\frac{2k_{\ell n}k_{\ell'n'}}{\Delta R}W_{\ell\ell'}^{nn'}&\simeq&\cos{[(k_{||}-k_{||}')R+\varphi_{k}\ell-\pi\ell/2-\varphi_{k'}\ell'+\pi\ell'/2]}\sinc{[(k_{||}-k_{||}')\Delta R/2]}\nonumber \\
&&-\cos{[(k_{||}+k_{||}')R+\varphi_{k}\ell-\pi\ell/2+\varphi_{k'}\ell'+\pi\ell'/2]}\sinc{[(k_{||}+k_{||}')\Delta R/2]}\, .
\end{eqnarray}
Note that $k_{||}=2\pi j/\Delta R$ produces
\begin{eqnarray}
\frac{2k_{\ell n}k_{\ell'n'}}{\Delta R}W_{\ell\ell'}^{nn'}&\simeq&\cos{[2\pi N(j-j')+\varphi_{k}\ell-\pi\ell/2-\varphi_{k'}\ell'+\pi\ell'/2]}\sinc{[\pi(j-j')]}\nonumber \\
&&-\cos{[2\pi N(j+j')+\varphi_{k}\ell-\pi\ell/2+\varphi_{k'}\ell'+\pi\ell'/2]}\sinc[\pi(j+j')] \nonumber \\
&=&\cos{[\varphi_{k}\ell-\pi\ell/2-\varphi_{k'}\ell'+\pi\ell'/2]}\sinc{[\pi(j-j')]}-\cos{[\varphi_{k}\ell-\pi\ell/2+\varphi_{k'}\ell'+\pi\ell'/2]}\sinc[\pi(j+j')] \nonumber \\
&=&\cos{[\varphi_{k}\ell-\pi\ell/2-\varphi_{k'}\ell'+\pi\ell'/2]}\delta_{j,j'} -\cos{[\varphi_{k}\ell-\pi\ell/2+\varphi_{k'}\ell'+\pi\ell'/2]}\delta_{j,-j'}\, .
\end{eqnarray}
\end{widetext}
This shows that the coupling $W_{\ell\ell'}^{nn'}$ makes modes with the same $k_{||}$ source each other in weak lensing for the SFB formalism, just as in the plane-parallel formalism.  Computationally, we have found that this conclusion is true for all $\ell$ and all but the first few of $k_{\ell n}$ which tend to couple with the first few $k_{\ell' n'}$'s, and we have found that neglecting these extra couplings do not affect our final results.  Thus, in our analysis while we perform the full integral in Eq.~\ref{E:wllnn}, we choose to only use the entries in the coupling matrix that correspond to $k_{||}=k_{||}'$, or $\sqrt{k_{\ell n}^2-(\ell/R)^2}=\sqrt{k_{\ell' n'}^2-(\ell'/R)^2}$ in order to dramatically reduce the computational effort.  Since this equality is never satisfied exactly, we instead find the $n'$ that most satisfies it for every $(\ell,n,\ell')$.  Also, when $\varphi_{k}\ell\simeq\varphi_{k'}\ell'$ one can use the asymptotic expansion to show that $\sqrt{\tau_{\ell n}\tau_{\ell'n'}}\sim2k_{\ell n}k_{\ell'n'}/\Delta R$, which motivates the approximation $W_{\ell\ell'}^{nn'}\simeq\delta_{\ell\ell'}^{nn'}/\sqrt{\tau_{\ell n}\tau_{\ell'n'}}$ where $\delta_{\ell\ell'}^{nn'}=1$ when the modes have the same $k_{||}$ and 0 otherwise.  This is only accurate at the 20\% level, so we do not use it for our main results; however, it is accurate for comparisons between different surveys where computing the full integrals would be very computationally expensive.}
%, and here we make a simple argument that these mode couplings are consistent with those for the plane-parallel approximation.  Based on the form of the integrand, $W_{\ell\ell'}^{nn'}$ in fact is largest when the first peaks of both spherical Bessel functions overlap in $r$, which occurs when $k_{\ell n}/\ell=k_{\ell'n'}/\ell'$.%  In the limit of small scales where the plane-parallel formalism should be accurate, $k\simeq k_\perp$ such that the equality condition becomes $k_\perp/\ell=k_\perp'/\ell'$.  This is manifestly true in the plane-parallel approximation since $k_\perp=\ell/R$.  This shows that the lensing coupling manifest in the plane-parallel lensing formalism will be preserved in our SFB formalism.
\newline

Using Eq.~\ref{E:theta}, it is possible to have the covariance matrix of the lensed signal for which we take the ensemble average over a fixed potential

{\begin{multline}
\langle\widetilde{T}_{\ell m n}\widetilde{T}_{\ell' m' n'}\rangle=\langle T_{\ell m n }T_{\ell' m' n'}\rangle\\ 
+\sum_{\ell_1 m_1}(-1)^{m_1}\left( {\begin{array}{ccc}
   \ell & \ell' & \ell_1\\
   m & m' & -m_1\\ 
   \end{array}} \right)\phi_{\ell_1m_1}f_{\ell\ell_1\ell'}^{nn'}\, ,
\label{eq:ttcov}
\end{multline}}
where $f_{\ell\ell_1\ell'}^{nn'}=M_{\ell\ell'}^{nn'}F_{\ell'\ell_1\ell}+M_{\ell'\ell}^{n'n}F_{\ell\ell_1\ell'}$, and $M$ is defined as
\begin{eqnarray}
M_{\ell\ell'}^{nn'}=\sum_{n''}\tau_{\ell n''}W_{\ell\ell'}^{n''n'}C_{\ell nn''}\, ,
\end{eqnarray}
{which can be written as an integral by using the integral forms of $W_{\ell\ell'}^{nn'}$ and $C_{\ell nn'}$ and the orthogonality condition from Eq.~\ref{E:jj2} to find
\begin{eqnarray}
M_{\ell\ell'}^{nn'}&=&\frac{2}{\pi}\int dk\,k^2P_T(k)[W_{\ell n}^0(k)-\beta W_{\ell n}^r(k)]\nonumber\\
&&\times[W_{\ell' n'\ell}^0(k)-\beta W_{\ell' n'\ell}^r(k)]\, ,
\end{eqnarray}
where
\begin{eqnarray}
W_{\ell n\ell'}^0=\int_{r_{\rm min}}^{r_{\rm max}} dr\,r^2j_\ell(k_{\ell n}r)j_{\ell'}(kr)\, ,
\end{eqnarray}
and
\begin{eqnarray}
W_{\ell n\ell'}^r=\int_{r_{\rm min}}^{r_{\rm max}} dr\,r^2j_\ell(k_{\ell n}r)j_{\ell'}''(kr)\, .
\end{eqnarray}
We also have the property $M_{\ell\ell}^{nn'}=C_{\ell nn'}$.}

{In the following work,} $C_{\ell nn'}$ is the unlensed {$T-T$} covariance, without any instrumental noise. {Also, here we clarify that we assume that the temperature maps have already had their beam and spectral window profiles deconvolved.} For the discussion that follows, note that the covariance with instrumental noise is denoted as
{\begin{eqnarray}
C^{\rm tot}_{\ell nn'}\equiv C_{\ell nn'}+C_{\ell nn'}^N\, ,
\end{eqnarray}}
{where $C_{\ell nn'}^N=N_{\ell nn'}/W_{\ell nn'}^{AB}$.} 

%Note that in Eq.~\ref{eq:ttcov}, the presence of a lensing potential makes the covariance acquire off-diagonal components in $\ell$, $m$ and $n$. This is due to two reasons- the breaking of isotropy by lensing and the fact that on the largest scales, {modes with different angular and radial values are coupled in the SFB formalism. On large angular scales, the $(k,\ell)$ modes can be oriented in nontrivial angles, whereas on small scales, the radial part of the mode can be described as $k_\parallel$, and any statistical invariance broken on the orthogonal plane keeps modes with different $k_\parallel$ uncorrelated.} However, on the largest scales, we find that broken isotropy naturally breaks homogeneity (radially) as well. This coupling is manifest in the formalism through $W_{\ell \ell'}^{nn'}$. Indeed, on small scales ($\ell\gg 1$), where the flat-limit is valid, $W_{\ell\ell'}^{nn'}\propto \delta_{n n'}$. The off-diagonal elements are what allow reconstruction of $\phi_{\ell m}$. Thus, we should expect lower noise on large scales than on small scales. In Section ~\ref{S:compare}, we see that this is the case. \newline

Now, we are prepared to consider a general quadratic estimator for the lensing potential. We use the following form, inspired by \cite{2003PhRvD..67h3002O},
{\begin{multline}
\hat{\phi}_{\ell m}=A_{\ell}\sum_{\ell_1 m_1 n_1}\sum_{\ell_2 m_2 n_2}(-1)^{m}\left( {\begin{array}{ccc}
   \ell_1 & \ell_2 & \ell\\
   m_1 & m_2 & -m\\ 
   \end{array}} \right) \\ 
   \times g_{\ell_1\ell_2\ell}^{n_1 n_2}\widetilde{T}_{\ell_1 m_1 n_1}\widetilde{T}_{\ell_2 m_2 n_2}\, ,
\label{eq:phihat}
\end{multline}}
where $A_\ell$ and $g_{\ell_1\ell_2\ell}^{n_1 n_2}$ are to be determined. {This is the most general quadratic estimator of the projected potential and we provide a proof of this fact in Appendix ~\ref{A:estimator}}. We require an unbiased estimator such that it produces the potential upon taking an ensemble average over many realizations of the 21 cm signal. To achieve this, we set the normalization condition

\begin{eqnarray}
\langle\hat{\phi}_{\ell m}\rangle|_{\text{lens}}=\phi_{\ell m}\, .
\end{eqnarray}
This condition straightforwardly produces the constraint

\begin{eqnarray}
\left(\frac{A_\ell}{2\ell+1}\right)^{-1}=\sum_{\ell_1\ell_2}\sum_{n_1n_2}g_{\ell_1\ell_2\ell}^{n_1n_2}f_{\ell_1\ell\ell_2}^{n_1n_2}\, .
\label{eq:Aell}
\end{eqnarray}
Finally, we require that the yet undetermined $g_{\ell_1\ell_2\ell}^{n_1n_2}$ minimize the variance of this estimator. To achieve this, we first compute the covariance of the estimator
{\begin{multline}
\langle\hat{\phi}_{\ell m}\hat{\phi}_{\ell' m'}^{*}\rangle=\delta_{\ell\ell'}\delta_{mm'}\frac{A_\ell A_{\ell'}^{*}}{2\ell+1}\sum_{\ell_1 n_1 n_1'}\sum_{\ell_2 n_2 n_2'}g_{\ell_1\ell_2\ell}^{n_1 n_2} \\ 
\times (g_{\ell_1\ell_2\ell}^{n_1' n_2'}{}^{*}C^{\rm tot}_{\ell_1 n_1 n_1'}C^{\rm tot}_{\ell_2 n_2 n_2'}+g_{\ell_2\ell_1\ell}^{n_1' n_2'}{}^{*}C^{\rm tot}_{\ell_1 n_1 n_2'}C^{\rm tot}_{\ell_2 n_2 n_1'})\, .
\end{multline}}
In order to minimize the variance, we require the functional derivative of $A_\ell$
\begin{eqnarray}
\frac{\delta A_\ell}{\delta g_{ab\ell}^{cd}}=\frac{-A_\ell^2}{2\ell+1}f_{a\ell c}^{c d}\, .
\end{eqnarray}
Minimising the variance thus produces
\begin{eqnarray}
S_{ab\ell}^{cd}=\frac{A_\ell}{2\ell+1}f_{a\ell c}^{cd}\sum_{\ell_1\ell_2}\sum_{n_1n_2}g_{\ell_1\ell_2\ell}^{n_1n_2}S_{\ell_1\ell_2\ell}^{n_1 n_2}\, ,
\label{eq:S}
\end{eqnarray}
where $S$ is defined to be
{\begin{eqnarray}
S_{ab\ell}^{cd}=\sum_{n_1n_2}[g_{ab\ell}^{n_1n_2}C^{\rm tot}_{acn_1}C^{\rm tot}_{bdn_2}+g_{ba\ell}^{n_1n_2}C^{\rm tot}_{acn_2}C^{\rm tot}_{bdn_1}]\, .
\end{eqnarray}}
One will notice that $g_{\ell_1\ell_2\ell}^{n_1n_2}$ is taken to be real in this equation. It follows straightforwardly from the parity properties of both $\phi_{\ell m}$ and {$T_{\ell m n}$}, and the properties of the 3j-symbol that in fact $g_{\ell_1\ell_2\ell}^{n_1n_2}{}^{*}=(-1)^{\ell_1+\ell_2+\ell}g_{\ell_1\ell_2\ell}^{n_1n_2}$. However, the presence of $F_{\ell_1\ell\ell_2}$ in Eq.~\ref{eq:S}, due to the properties of the 3j-symbol, forces the reality of $g_{\ell_1\ell_2\ell}^{n_1n_2}$, so we make this change early in hindsight for convenience. With this in mind, we find that Eq.~\ref{eq:S}, through the definition of $A_\ell$ in Eq.~\ref{eq:Aell}, leads to the equality
{\begin{eqnarray}
f_{a\ell c}^{cd}&=&S_{ab\ell}^{cd}\nonumber \\
&=&\sum_{n_1n_2}[g_{ab\ell}^{n_1n_2}C^{\rm tot}_{acn_1}C^{\rm tot}_{bdn_2}+g_{ba\ell}^{n_1n_2}C^{\rm tot}_{acn_2}C^{\rm tot}_{bdn_1}]\, .
\end{eqnarray}}
Using (column permutation) properties of the 3j-symbol once again, one can derive that $g_{ab\ell}^{n_1n_2}=(-1)^{a+b+\ell}g_{ba\ell}^{n_2 n_1}$, where the phase factor is again omitted in hindsight for the purpose of convenience. Thus, we get
{\begin{eqnarray}
\sum_{n_1 n_2}g_{ab\ell}^{n_1n_2}C^{\rm tot}_{a c n_1}C^{\rm tot}_{bdn_2}=\frac{f_{a\ell c}^{cd}}{2}\, .
\end{eqnarray}}
For convenience, we temporarily define {$\xi_{ab\ell}^{n_1 d}$} as
{
\begin{eqnarray}
\xi_{ab\ell}^{n_1 d}=\sum_{n_2}g_{cd\ell}^{n_1 n_2}C^{\rm tot}_{bdn_2}\, .
\end{eqnarray}
}
We solve for {$\xi$} first, the equation for which is now
{
\begin{eqnarray}
\sum_{n_1}\xi_{ab\ell}^{n_1d}C^{\rm tot}_{acn_1}=\frac{f_{a\ell c}^{cd}}{2}\, .
\end{eqnarray}
}
Now, this is linear so we simply invert it. To be specific, for fixed $a$, $(C^N_{a}{}^{-1})_{nn'}$ is the $(n,n')$ element of the inverse of the (symmetric) matrix $C_{a}$, i.e., the $N\times N$ matrix with elements $(C_a)_{nn'}=C_{ann'}$. Solving this now presents
{
\begin{eqnarray}
\xi_{ab\ell}^{cd}=\sum_{n_1}(C_{a}^{\rm tot}{}^{-1})_{cn_1}\frac{f_{a\ell c}^{n_1d}}{2}\, .
\end{eqnarray}
}
Proceeding identically as above to solve to solve for $g$ presents
{\begin{eqnarray}
g_{ab\ell}^{cd}=\frac{1}{2}\sum_{n_1n_2}(C_{a}^{\rm tot}{}^{-1})_{cn_1}(C_{b}^{\rm tot}{}^{-1})_{dn_2}f_{a\ell b}^{n_1n_2}\, .
\end{eqnarray}}
This completes our derivation of the minimum variance {weighting for} the lensing potential. Further, we have for the lensing potential
\begin{eqnarray}
\langle\hat{\phi}_{\ell m}\hat{\phi}_{\ell' m'}^{*}\rangle=\delta_{\ell\ell'}\delta_{mm'}(C_{\ell}^{\phi\phi}+N_{\ell}^{\phi\phi})\, .
\end{eqnarray}
We assume a fiducial survey, so we have for the noise power spectrum of the lensing potential the expression
{\begin{eqnarray}
\left(\frac{N_\ell^{\phi\phi}}{2\ell+1}\right)^{-1}&=&\left(\frac{\langle\norm{\hat{\phi}_{\ell m}}^2\rangle}{2\ell+1}\right)^{-1}\nonumber \\
&=&\frac{1}{2}\sum_{\ell_1\ell_2}\sum_{n_1n_2}\sum_{n_1'n_2'}(C_{\ell_1}^{\rm tot}{}^{-1})_{n_1n_1'}(C_{\ell_2}^{\rm tot}{}^{-1})_{n_2n_2'}\nonumber \\
&& \times f_{\ell_1\ell\ell_2}^{n_1'n_2'}f_{\ell_1\ell\ell_2}^{n_1n_2}\, .
\end{eqnarray}}
Additionally, we derive the flat-sky limit of our estimator in order to draw a rough comparison with the estimator of \cite{2006ApJ...653..922Z}. The flat-sky noise power spectrum is:

{\begin{eqnarray}
N^{\phi\phi}(\ell)^{-1}&=&\frac{1}{2}\sum_{n_1 n_2}\sum_{n_1'n_2'}\int\frac{d^2\ell_1}{(2\pi)^2}(C_{\ell_1}^{\rm tot}{}^{-1})_{n_1n_1'}(C_{\ell_2}^{\rm tot}{}^{-1})_{n_2n_2'}\nonumber\\
&&\times \bar{f}_{\ell_1\ell\ell_2}^{n_1'n_2'}\bar{f}_{\ell_1\ell\ell_2}^{n_1n_2}\, ,
\end{eqnarray}
where $\bar{f}_{\ell_1\ell\ell_2}^{n_1 n_2} = M_{\ell_1 \ell_2}^{n_1 n_2} \boldsymbol{\ell}\cdot\boldsymbol{\ell_1}+M_{\ell_2 \ell_1}^{n_2 n_1} \boldsymbol{\ell}\cdot\boldsymbol{\ell_2}$ and $\boldsymbol{\ell_2}=\boldsymbol{\ell}-\boldsymbol{\ell_1}$.}
In Appendix \ref{app:A}, we present a full derivation of the flat-sky limit of our estimator and the above expression.

{We perform our forecast by approximating the covariance matrix as diagonal, $C_{\ell n n'}^{\rm tot}=C_{\ell n}^{\rm tot}\delta_{nn'}$.  This is justified since for most $k_{\ell n}$, $W_{\ell n}(k)$ and $W^r_{\ell n}(k)$) are highly peaked and the $k_{\ell n}$ values are separated enough to have little window function overlap.  For $k_{\ell n}R\sim \ell$, we find that while $C_{\ell nn'}$ is still diagonal in the non-RSD case ($\lvert C_{\ell nn'}\rvert< 10\%$ of $C_{\ell nn}$), we see that RSD can cause the off-diagonal terms to be significant. However, we expect that this should not affect the overall behavior of our results. We leave this investigation for future work. Considering the diagonal sets the quantity $M_{\ell\ell'}^{nn'}=\tau_{\ell n}W_{\ell\ell'}^{nn'}C_{\ell n}$ while trivially rewriting the sum
\begin{eqnarray}
\sum_{n_1 n_2}\sum_{n_1'n_2'}(C_{\ell_1}^{\rm tot}{}^{-1})_{n_1n_1'}(C_{\ell_2}^{\rm tot}{}^{-1})_{n_2n_2'}f_{\ell_1\ell\ell_2}^{n_1'n_2'}f_{\ell_1\ell\ell_2}^{n_1n_2}\nonumber\\
=\sum_{n_1n_2}\frac{\left(f_{\ell_1\ell\ell_2}^{n_1n_2}\right)^2}{C_{\ell_1 n_1}^{\rm tot}C_{\ell_2 n_2}^{\rm tot}}\, .
\end{eqnarray}
Note that applying the approximation $\tau_{\ell n}C_{\ell n}\simeq R^2\Delta R C_\ell(k_{||,j})$ to the SFB flat-sky expression for $N_\ell^{\phi\phi}$ will give you the plane-parallel expression in Eq.~\ref{E:nlpp}.  This suggests that the plane parallel and SFB (flat and full sky) formalisms will deviate when this approximation breaks down.}

%We already know the approximation breaks down for $\ell > k_{\rm eq}R$.  In particular, we expect the sums in $N_\ell^{\phi\phi}$ to increase for $\ell_1 > k_{\rm eq}R$ and $\ell_2>\ell_1$ since $C_{\ell n}$ decreases with $\ell$.  In addition, we expect that the sum in the lensing noise power spectrum $N_\ell^{\phi\phi}$ will terminate when the instrumental noise power spectrum $C_{\ell n}^N$ starts to increase due to finite angular resolution.  

{We already know the approximation breaks down for $\ell \sim k_{\ell n}R$.  If the approximation were always valid, then in the low instrumental noise regime the sums in $N_\ell^{\phi\phi}$ would stays constant at low $\ell$ where the $k_{\ell n}$'s are approximately $\ell$-independent and would increase when the $k_{\ell n}$'s are approximately $\ell$-dependent around $\ell=2\pi R/\Delta R$.  This is what occurs in the plane-parallel formalism and gives the same result, in particular a constant $N_\ell^{\phi\phi}$ at low $\ell$.  However, because the approximation breaks down, at $k_{\ell n}\sim \ell$ the situation changes where the autocorrelation of modes with the same value of $k_{\ell n}$ but different $\ell$'s differ, causing the sums in $N_\ell^{\phi\phi}$ to increase for $\ell_1 > k_{\rm eq}R$ and $\ell_2>\ell_1$ where $C_{\ell n}$ decreases with $\ell$.  In addition, we expect that the sum in the lensing noise power spectrum $N_\ell^{\phi\phi}$ will terminate when the instrumental noise power spectrum $C_{\ell n}^N$ starts to increase due to finite angular resolution.  Thus, we expect the SFB prediction for $N_\ell^{\phi\phi}$ to deviate from the plane-parallel prediction when there are a significant number of angular modes $k_{\rm eq}R<\ell<\ell_{\rm res}$, where modes $\ell>\ell_{\rm res}$ are beam smeared.  In particular, the modes $T(\boldsymbol{\ell}_2)$ that are correlated with $T(\boldsymbol{\ell}_1)$ due to  $\phi(\boldsymbol{L})$ are constrained by the range $|\boldsymbol{L}-\boldsymbol{\ell}_1|<\boldsymbol{\ell}_2<\boldsymbol{L}+\boldsymbol{\ell}_1$ such that for small enough $\boldsymbol{L}$, $\boldsymbol{\ell}_1$ and $\boldsymbol{\ell}_2$ will be close enough together and in the right range so that both $\tau_{\ell_1n_1}C_{\ell_1n_1}$ and $\tau_{\ell_2n_2}C_{\ell_2n_2}$ will deviate from $R^2\Delta R C_\ell(k_{||,j})$ while still being contributing to the sum.  Specifically, we set $L_{\rm dev}$ as the scale at which for $L<L_{\rm dev}$ the SFB and plane-parallel predictions for $N_\ell^{\phi\phi}$ deviate, and we define it by setting the condition $\boldsymbol{\ell}_{\rm 2,max}=\boldsymbol{L}+\boldsymbol{\ell}_1<\ell_{\rm res}\forall\,L<L_{\rm dev}$ when $\boldsymbol{\ell}_1>k_{\rm eq}R$.  Simplifying these conditions gives us the deviation scale
\begin{eqnarray}
L_{\rm dev}=\ell_{\rm res}-k_{\rm eq}R\, ,
\end{eqnarray}
and this equation seems to work well when we set $\ell_{\rm res}$ to be the angular mode where the instrumental noise is 25\% higher than the value at $\ell=0$ (see Sec.~\ref{S:compare}).  Note that this is correspondingly where the terms for the sums in $N_\ell^{\phi\phi}$ decrease by 50\%. As a physical interpretation, we expect that for low angular resolution (low $\ell_{\rm res}$) or low curvature (high $R$), we cannot distinguish modes from different directions in harmonic space, such that the lensing noise reduces to the plane-plane parallel case.}

\section{Comparison between estimators}\label{S:compare}

\begin{figure}
\begin{center}
\includegraphics[width=0.55\textwidth]{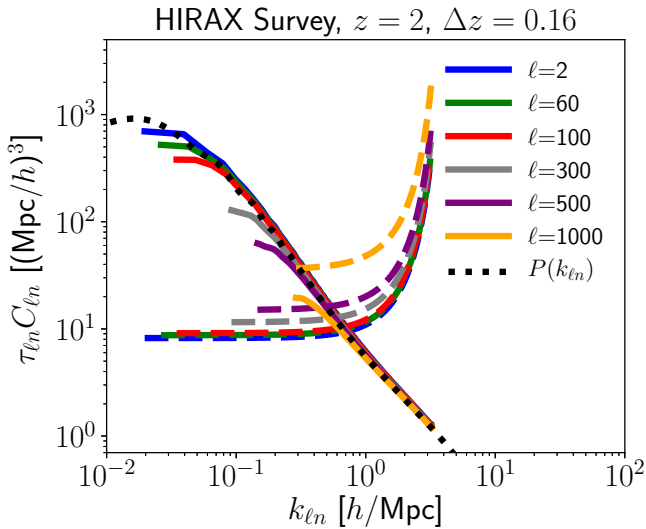}
\caption{\label{F:cln}  {The SFB power spectra for signal (solid) and noise (dashed) in the range $\ell=2-1000$.  The $\ell=2-100$ noise curves are barely distinguishable because they overlap with each other.  We also include a black,dotted curve for $P(k_{\ell n})$ including RSD where we set $\mu=1$}  The signal applies a redshift $z=2$ with a window $\Delta z=0.16$ and {changes negligibly as $\ell$ varies except at $k_{\ell n}R\sim\ell$.}  The noise applies the properties of the HIRAX 21-cm survey.  {The minimum $k$ for each curve is set by the SFB series; it is not due to foreground or nonlinear cuts.}}
\end{center}
\end{figure}

Here we compare our full-sky and flat-sky expressions for $N_\ell^{\phi\phi}$ to that from \citet{2006ApJ...653..922Z}.  Similar to the forecast in \citet{2018JCAP...07..046F}, hereafter F18, we assume the survey {to be over the redshift range $1.38\leq z\leq 2.57$.}  For the 21-cm signal, we assume the same form used in F18 relevant at low-redshifts given as
\begin{eqnarray}
T(z)=0.3\left(\frac{\Omega_{\rm HI}}{10^{-3}}\right)\left(\frac{\Omega_m+(1+z)^{-3}\Omega_\Lambda}{0.29}\right)^{-1/2}\left(\frac{1+z}{2.5}\right)^{1/2}{\rm mK}\, ,
\end{eqnarray}
where $\Omega_{\rm HI}=5\times10^{-4}$ \citep{2013ApJ...763L..20M}.  We write the power spectrum ${P_T(k,z)=b_{\rm HI}^2(z)T^2(z)P(k,z)}$, where ${P(k,z)}$ is the matter power spectrum computed from CAMB \citep{2000ApJ...538..473L} with the nonlinear clustering regime computed from HALOFIT \citep{2003MNRAS.341.1311S}, and $b_{\rm HI}$ is the clustering bias of HI emitters set {based on predictions from \citet{2017MNRAS.471.1788C} to the values in Table \ref{T:bias}.}  Note that we also include the anisotropy from redshift-space distortions (RSD), specifically the Kaiser effect.  For the SFB power spectra, the Kaiser effect is given explicitly in Eq.~\ref{E:clnn}, while for the plane-parallel power spectra this is accounted by adding an extra factor $(1+\beta\mu^2)^2$ to the power spectrum $P(k)$ {where $\beta=f_g/b_{\rm HI}$ and the growth rate $f_g(z)=[\Omega_m(z)]^{0.55}$}.  The SFB power spectrum for the signal is shown in Fig.~\ref{F:cln}.
\begin{table}
    \centering
    \begin{tabular}{|c||c}
        \cline{1-2}
          z & $b_{\rm HI}$\\ \hline\hline
          $1.38\leq z\leq 1.48$ & 1.6  \\ \hline
          $1.48\leq z\leq 1.59$ & 1.7   \\ \hline
          $1.59\leq z\leq 1.71$ & 1.8   \\ \hline 
          $1.71\leq z\leq 1.84$ & 1.9   \\ \hline
          $1.84\leq z\leq 1.99$ & 2.0   \\ \hline
          $1.99\leq z\leq 2.16$ & 2.0   \\ \hline
          $2.16\leq z\leq 2.35$ & 2.2  \\ \hline
          $2.35\leq z\leq 2.57$ & 2.4 \\ \hline\hline
    \end{tabular}
    \caption{{HI clustering bias values used for our forecasts based on predictions from \citet{2017MNRAS.471.1788C}.}}
    \label{T:bias}
\end{table}

We assume the noise properties of the HIRAX survey \citep{2016SPIE.9906E..5XN} given in \citet{2018JCAP...07..046F} to forecast the noise power spectra.  {In particular, we assume the survey covers half the sky with an angular resolution of 11 arcmin at $z=2$.}  For the plane-parallel formalism, the noise power spectrum for HIRAX is given by
\begin{eqnarray}
C_\ell^N(k_\parallel)=\frac{T_{\rm sys}^2(\nu)}{t_{\rm pix}B}\frac{A_{\rm pix}}{W(\ell)}\, ,
\end{eqnarray}
where $T_{\rm sys}(\nu)$ is instrumental thermal noise, $B$ is the bandwidth corresponding to the redshift window, $A_{\rm pix}$ and $t_{\rm pix}$ are the angular area and observing time per angular pixel, and $W(\ell)$ is the beam window function.  For all these quantities we use the same values as in F18.  For the SFB noise power spectrum, we set {$C_{\ell n}^N$} in terms of $C_\ell^N$ based on Eq.~\ref{E:Nln} as
{\begin{eqnarray}
C_{\ell n}^N=\frac{C_\ell^N}{\tau_{\ell n}W_{\ell nn}^A}R^2\Delta R\, .
\end{eqnarray}}
Using this formalism, we plot {$C_{\ell n}$ and $C_{\ell n}^N$} in Fig.~\ref{F:cln}.  We see that for the SFB formalism, the signal is dominant over the noise up to {$k_{\ell n}\simeq 0.7$ $h$/Mpc when the noise begins to exceed the signal.  In particular, the beam window function eliminates scales greater than around $\ell=1000$ and the spectral window function starts to matter around $k_{\ell n}\simeq 2$ $h$/Mpc.}

{In our $N_\ell^{\phi\phi}$ forecasts we consider cuts to avoid the contamination from continuum foregrounds.  In the plane-parallel (PP) basis for a single-dish survey the contamination is confined to low-$k_\parallel$ modes.  It was shown in L17 that in the SFB basis low-$k$ modes are contaminated over the range $\sim1/\Delta R$ except at large $\ell$ where the contamination reaches out to higher $\ell$.  In Appendix~\ref{A:fore} we explain this behavior in that the contours of constant contamination are equivalent to contours of constant $k_{||}=\sqrt{k^2-(\ell/R)^2}$.  Thus in order to simulate equivalent foreground cuts for both the plane-parallel and SFB formalisms, we set $k_{\rm min} = \sqrt{k_{\rm ||,min}^2+(\ell/R)^2}$, where $k_{\rm ||,min}=2\pi j_{\rm min}/\Delta R$ and $j_{\rm min}$ is the minimum $j$ allowed by foreground cuts in the PP result.}

{We also simulate the removal of nonlinear modes by setting $k_{\rm max}=0.5 h$/Mpc.}  Note that we do not use bias-hardened estimators in either formalism as introduced in F18, which will be needed to remove correlations due to nonlinear clustering.  We leave this for future work.

\begin{figure}
\begin{center}
\includegraphics[width=0.5\textwidth]{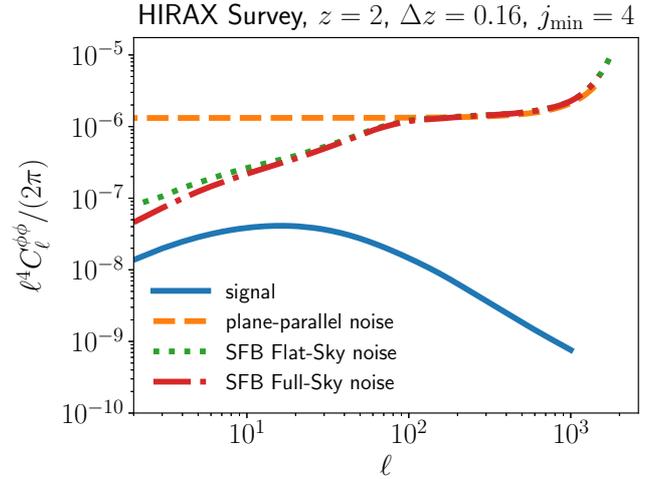}
\caption{\label{F:Nlpp}  Our forecasts for the lensing noise bias for the PP formalism of \citet{2006ApJ...653..922Z} (dashed) along with our SFB flat sky (dotted) and full sky (dash-dotted).  We also include a $C_\ell^{\phi\phi}$ prediction (solid).  These forecasts apply the properties of the HIRAX 21-cm survey {with limits of $j_{\rm min}=4$ and $k_{\rm max}\leq0.5h$/Mpc to limit contamination from foregrounds and nonlinear modes, respectively.}  Note that for scales {$\ell\lesssim 144$} our SFB predictions deviate from the plane-parallel predictions by as much as a factor of {26}.}
\end{center}
\end{figure}

We plot forecasts for $N_\ell^{\phi\phi}$ for the full-sky SFB formalism in Fig.~\ref{F:Nlpp} along with forecasts assuming the flat-sky SFB formalism we derived and the flat-sky, {PP} formalism from \citet{2006ApJ...653..922Z}.  {We simulate reasonable foreground cuts by setting $j_{\rm min}=4$.}  We see that the PP and SFB forecasts agree at small scales, but deviate dramatically at {$\ell_{\rm dev}\simeq 144$} and differing by as much as a factor of {26} at $\ell=2$. This would imply that $C_\ell^{\phi\phi}$ at the largest scales could be more accessible than previously thought.  {Note that with R(z=2)=3594 Mpc/$h$ and $k_{\rm eq}=0.02h$/Mpc for our fiducial cosmology and $\ell_{\rm res}=205$, we would guess a value $\ell_{\rm dev}=\ell_{\rm res}-k_{\rm eq}R=133$ which is close to where the plane-parallel and SFB curves deviate in Fig.~\ref{F:Nlpp}.}  We also see that the deviation between the SFB flat and full sky cases is small  {and takes place for $\ell\lesssim10$} with a magnitude similar to the CMB flat and full sky cases.  {The SFB full-sky vs flat-sky split location $\ell\sim10$ is a result of the difference between $F_{\ell_1L\ell_2}$  and $\boldsymbol{L}\cdot\boldsymbol{\ell_1}$ which leads to a fractional difference for $N_\ell^{\phi\phi}$ of $1/(L+1)^2$, implying a >1\% difference for $\ell\lesssim10$.}% This implies that the major cause of the plane-parallel failure at large angular scales is not the sky curvature directly, but the sky curvature causing both the $k$ and $\ell$ modes to be lensed.  This greatly allows the intensity perturbations to be more sensitive to $\phi$ than in the plane-parallel case where only the $\ell$ modes are lensed and not the $k_\parallel$ modes.
%\begin{figure}
%\begin{center}
%\includegraphics[width=0.5\textwidth]{figs/Nl_phi_hirax_diff.eps}
%\caption{\label{F:Nlppdiff}  The difference in the noise bias between the SFB full-sky formalism and the 2 alternative formalisms, relative to the full sky prediction. \textbf{Will replace with SFB predictions with k cutoffs instead of window functions.}}
%\end{center}
%\end{figure}

\subsection{Varying survey conditions}

\begin{figure}
\begin{center}
\includegraphics[width=0.5\textwidth]{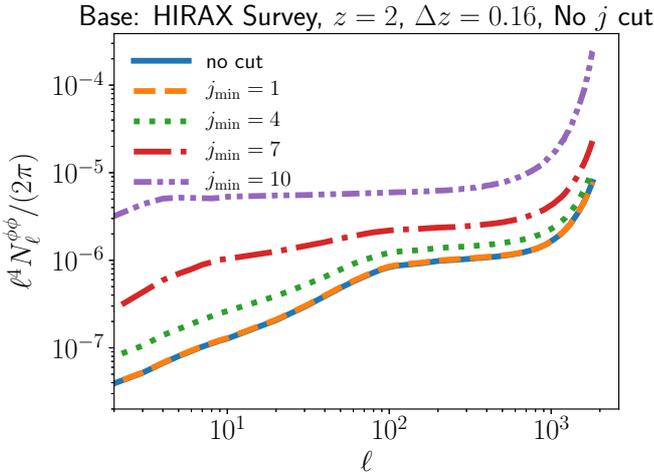}
\caption{\label{F:Nlppfore}  {$N_\ell^{\phi\phi}$ for various levels of foreground cuts.  Note that the the curve for $j_{\rm min}=1$ agrees with the curve without any cut within 2\%.  Low levels of foreground removal will preserve the decrease in $N_\ell^{\phi\phi}$ at low $\ell$, while higher levels of cuts will cause $N_\ell^{\phi\phi}$ to be less steep at low $\ell$.}}
\end{center}
\end{figure}

{In this subsection we consider the effects of varying foreground cuts, the survey shell volume, and the angular resolution of the survey. Note that in all our examples we only compare $N_\ell^{\phi\phi}$ curves using SFB flat-sky instead of SFB full-sky since they are equal in most of the $\ell$-range and SFB flat-sky is much faster to compute.  Also, based on our argument in the previous section we do not expect the SFB full-sky vs flat-sky split location $\ell\sim10$ to vary with survey parameters. First, we consider how $N_\ell^{\phi\phi}$ is affected by the level of foreground cuts.  We plot the SFB flat-sky forecasts with no foreground cut and with various values for $j_{\rm min}$ between 1 and 10 in Fig.~\ref{F:Nlppfore}.}

{First we find that although some modes are excised for $j_{\rm min}=1$, $N_\ell^{\phi\phi}$ is not affected by this cut ($\lesssim2$\% difference).  Next, we find we would gain at most a 50\% reduction in the noise by applying less stringent cuts.  Finally, we see that in the limit of very stringent cuts, $j_{\rm min}\sim 10$, all the modes with $k_{\ell n}\sim\ell/R$ are removed such that $k_{\ell n}\gg\ell/R$ which is equivalent to $k_{\ell n}\sim k_{||}$ and that the result approaches a form similar to PP. It appears that $j_{\rm min}$ plays a role not in altering the behavior of $L_{\rm dev}$ but in fact the slope of $N_{\ell}^{\phi\phi}$ around $L_{\rm dev}$. To see why this is the case, note that the summand for $N_{\ell}^{\phi\phi}$ is roughly constant for low $\ell$s but is strongly suppressed at high $\ell$s ($\ell>\ell_{\rm res}$). While this suppression is immune to the choice of $j_{\rm min}$, the low $\ell$ behavior is suppressed by the increase of $j_{\rm min}$. Thus increasing $j_{\rm min}$ naturally produces a lower slope.
Most importantly, we find that the reduction in the noise at low $\ell$ is present for reasonable levels of foreground cutting and will be prudent to consider.}

\begin{figure}
\begin{center}
\includegraphics[width=0.5\textwidth]{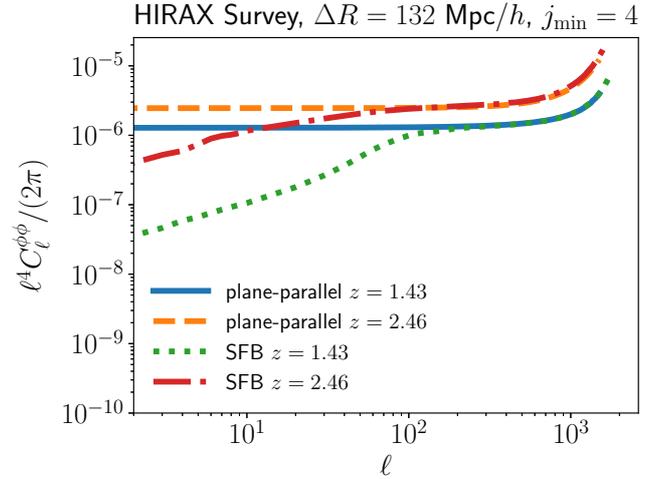}
\caption{\label{F:Nlppr}  {SFB and PP forecasts for $N_\ell^{\phi\phi}$ at redshifts $z=1.43$ and $z=2.46$. The comoving distances (shell radii) corresponding to these redshifts are $R=2950$ and 4008 Mpc/$h$, respectively.  For these cases we also set $\Delta z=0.1$ and 0.162 so that $\Delta R$ for these redshifts would be equal.  The $\ell_{\rm res}$ values for these redshifts being 241 and 185, respectively, would predict $\ell_{\rm dev}=182$ and 105, which are close to the apparent locations.  We also see an increase in the overall noise from $z=1.43$ to $z=2.46$ due to the latter having a higher $T_{\rm sys}$.}}
\end{center}
\end{figure}

\begin{figure}
\begin{center}
\includegraphics[width=0.5\textwidth]{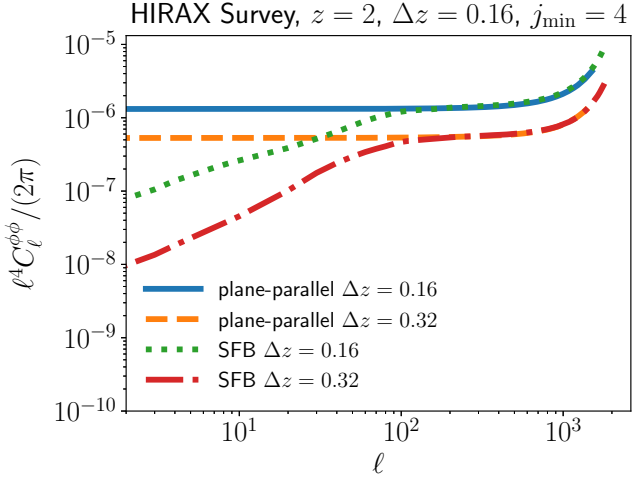}
\caption{\label{F:Nlppdz}  {SFB and PP forecasts for $N_\ell^{\phi\phi}$ at redshift $z=2$ for $\Delta z=0.16$ and 0.32. $\ell_{\rm res}$ is independent of $\Delta R$, which would imply that $\ell_{\rm dev}$ for both curves should be the same, which is what we find numerically.  Note that the noise curve for $dz=0.32$ is lower because the bandwidth of the survey is longer.}}
\end{center}
\end{figure}

{Next we consider how $N_\ell^{\phi,\phi}$ changes if we change the survey volume redshift parameters $z$ and $\Delta z$.  Note that we reduce the computational effort by using the approximation $W_{\ell\ell'}^{nn'}\simeq\delta_{\ell\ell'}^{nn'}/\sqrt{\tau_{\ell n}\tau_{\ell'n'}}$ for all $N_\ell^{\phi\phi}$ curves.  In Fig.~\ref{F:Nlppr} we plot the PP and SFB $N_\ell^{\phi\phi}$ curves for redshifts $z=1.43$ and $z=2.46$. Based on the corresponding values of $\ell_{\rm res}$ and $R$ (see Fig.~\ref{F:Nlppr} for actual values), our formula would predict $\ell_{\rm dev}=182$ and 105, which are close to the apparent locations in the figure. Note that the curve for $z=2.46$ is lower because $T_{\rm sys}$ is higher leading to higher $C_\ell^N$.  In Fig.~\ref{F:Nlppdz} we plot the PP and SFB $N_\ell^{\phi\phi}$ curves for redshifts $z=2$ with both $\Delta z=0.16$ and 0.32.  In the figure, $\ell_{\rm dev}$ is the same for both curves, which is consistent with our model since $\ell_{\rm res}$ is independent of $\Delta R$.  We do see the noise decrease for $\Delta z=0.32$ since the bandwidth $B$ is longer.}

\begin{figure}
\begin{center}
\includegraphics[width=0.5\textwidth]{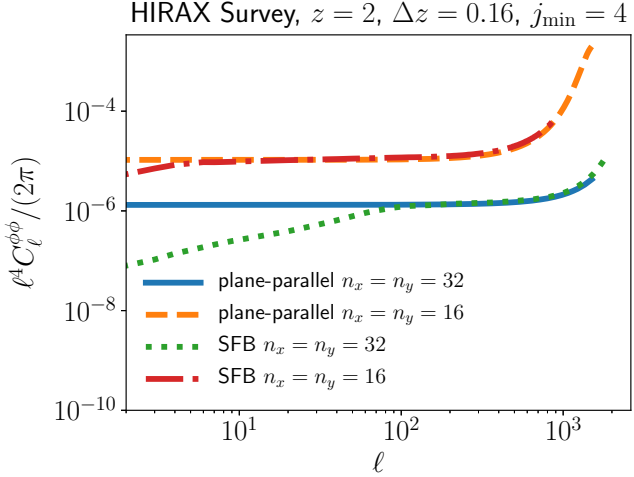}
\caption{\label{F:Nlppnx}  {SFB and PP forecasts for $N_\ell^{\phi\phi}$ at redshift $z=2$ ($R=3594$ Mpc/$h$) for higher ($n_x=n_y=32$) and lower ($n_x=n_y=16$) angular resolution. The $\ell_{\rm res}$ values for these cases being 205 and 99, respectively, would predict $\ell_{\rm dev}= 134$ and 28.  The prediction for the lower resolution curve is a bit higher than the apparent location ($\ell\sim15$), which is of the right order but may show a limitation of our $\ell_{\rm dev}$ formula.  Note that the noise curve for the lower angular resolution case is higher since the higher number of beams increases $C_\ell^N$.}}
\end{center}
\end{figure}

{Finally we consider what happens when we reduce the angular resolution in Fig.~\ref{F:Nlppnx}.  We model this for the HIRAX survey by reducing the number of dishes in both orthogonal directions of the array ($n_x$ and $n_y$) by half from 32 to 16. Based on the corresponding values of $\ell_{\rm res}$ (see Fig.~\ref{F:Nlppnx} for actual values), our formula would predict $\ell_{\rm dev}= 134$ and 28 for the higher and lower resolution cases, respectively.  The predicted $\ell_{\rm dev}$ value for the lower resolution case is a bit higher than the apparent location in the figure, which is more like $\ell\sim15$. It is not obvious what is causing the difference; it could be that for a changing angular resolution our rule of a 25\% increase in $C_{\ell n}^N$ is not adequate, but it is more likely that it just appears worse than the other examples because $\ell_{\rm dev}$ is so low in this case.  Either way, the result is still of the right order, and it still confirms the directional behavior we predicted in our $\ell_{\rm dev}$ model.  We also see that the overall noise for lower angular resolution is higher because the total number of beams $n_{\rm beam}\sim 4n_xn_y$ is inversely proportional to the noise $C_\ell^N$.}

%We also consider how $N_\ell^{\phi\phi}$ changes with respect to changes in the survey parameters.  The parameters we consider include $z$, $\Delta z$, and $N_{\ell n}$.  We show specific results in Fig.~\ref{F:Nlppalts}.  We find that the flat part of the spectrum is proportional to $(\Delta R/R)C_\ell^N$, while the $\ell$ where the spectrum decreases does not change.  We also consider neglecting to remove foreground-contaminated, large-radial-scale modes from the estimator.  We find that this barely affects the noise bias, though obviously this would greatly increase the foreground contamination.  {We also predict that the noise at the lower redshift $z=1$ will be about 10$\times$ less than for $z=2$, although we did not include nonlinear clustering in our fiducial power spectrum so we expect this to be mitigated.  Note that nonlinear clustering will affect predictions of the noise bias at all redshifts.}
%\begin{figure}
%begin{center}
%\includegraphics[width=0.5\textwidth]{figs/Nl_phi_hirax_alts.eps}
%\caption{\label{F:Nlppalts}  $N_\ell^{\phi\phi}$ under various configurations, including the base case (solid), changing $\Delta z=0.32$ (dashed), changing $z=1$ (dotted), neglecting foreground removal (dot-dashed), and improving the noise directly by a factor of 10 (dot-dot-dashed).}
%\end{center}
%\end{figure}

\subsection{Effects to lensing power spectrum measurements}

With our prediction for $N_\ell^{\phi\phi}$, we can now forecast the uncertainty on $C_\ell^{\phi\phi}$, given by
\begin{eqnarray}
\sigma(C_\ell^{\phi\phi})=\sqrt{\frac{2}{(2\ell+1)f_{\rm sky}}}N_\ell^{\phi\phi}\, ,
\end{eqnarray}
where $f_{\rm sky}=0.5$ is the assumed sky fraction mapped by the HIRAX survey and we take the limit $C_\ell^{\phi\phi}\ll N_\ell^{\phi\phi}$. We find that the uncertainty for $C_\ell^{\phi\phi}$ in the SFB formalism decreases significantly in comparison to the plane-parallel case at scales where spherical geometry becomes relevant. In Table~\ref{tab:snr_pp}, we list the SNR values of $\phi\phi$ autocorrelation for the Plane-Parallel case and our SFB case {over the redshift range $1.38\leq z\leq 2.57$, which corresponds to the lower half of the HIRAX frequency range}.
Particularly, we find an increase of {107\%} for the SNR.  We note, however, that the SNR values are {likely} overestimated due to the fact that we have not used bias-hardened estimators for this calculation, {as well as that nonlinear clustering was not included in the fiducial power spectrum.  We expect to derive relevant results for a bias-hardened full-sky estimator for $\Phi$ in future work.}  
\begin{figure}
\begin{center}
\includegraphics[width=0.5\textwidth]{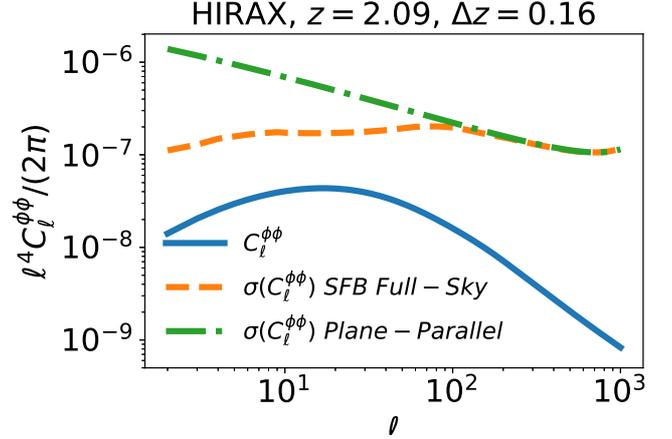}
\caption{\label{F:clpp}  The forecasted signal for $C_\ell^{\phi\phi}$ along with errors for both the plane-parallel estimator with the SFB full-sky estimator.  We find that the noise using the full-sky estimator decreases dramatically at large scales.}
\end{center}
\end{figure}

Following the example in F18, we also consider the uncertainty on the angular cross-power spectrum $C_\ell^{\kappa g}$, where $\kappa=-\nabla^2\phi/2$
is the lensing convergence and $g$ represents a galaxy distribution.  For cross-correlations, $\kappa$ is generally favored since it directly traces the matter perturbations.  The expression for $C_\ell^{\kappa g}$ is given by
{\begin{eqnarray}
C_\ell^{\kappa g}=\frac{3}{2c^2}\Omega_{m,0}H_0^2\frac{2}{\pi}\int dk\,k^2W_\ell^\kappa(k)W_\ell^g(k)P(k)\, ,
\end{eqnarray}
where
\begin{eqnarray}
W_\ell^\kappa(k)&=&\int_0^{\chi_s}d\chi\,\chi\frac{\chi_s-\chi}{\chi_s}(1+z)D(z)j_\ell(k\chi)\nonumber\\
W_\ell^g(k)&=&\int dz\,b_gf(z)D(z)j_\ell(k\chi)\, ,
\end{eqnarray}}
%\begin{eqnarray}
%C_\ell^{\kappa g}=\frac{3}{2c^2}\Omega_{m,0}H_0^2\int_0^{z_s}dz\frac{\chi_s-\chi}{\chi_s\chi}(1+z)f(z)b_gP\left(\frac{\ell}{\chi},z\right)\, .
%\end{eqnarray}
$b_g$ is the galaxy clustering bias, and $f(z)$ is the redshift distribution.  The uncertainty on $C_\ell^{\kappa g}$ is given by
\begin{eqnarray}
\sigma^2(C_\ell^{\kappa g})=\frac{1}{(2\ell+1)f_{\rm sky}^{\kappa g}}\left[(C_\ell^{\kappa g})^2+N_\ell^{\kappa\kappa}(C_\ell^{gg}+N^{gg})\right]\, ,
\end{eqnarray}
where $N_\ell^{\kappa\kappa}=[\ell(\ell+1)/2]^2N_\ell^{\phi\phi}$, $C_\ell^{gg}$ is the galaxy angular auto-power spectrum given by
{\begin{eqnarray}
C_\ell^{gg}=\frac{2}{\pi}\int dk\,k^2[W_\ell^g(k)]^2P(k)\, ,
\end{eqnarray}}
%\begin{eqnarray}
%C_\ell^{gg}=\int dz\, \frac{H(z)}{c}\frac{f^2(z)}{\chi^2}b_g^2P\left(\frac{\ell}{\chi},z\right)\, ,
%\end{eqnarray}
$N^{gg}=1/\overline{n}_g$ is the galaxy shot noise related to the areal number density $\overline{n}_g$, $f_{\rm sky}^{\kappa g}$ is the sky fraction observed by both the {LIM} lensing and galaxy surveys, and we take the limit $C_\ell^{\kappa\kappa}\ll N_\ell^{\kappa\kappa}$.  As in F18, we assume the specifications from the Large Synoptic Survey Telescope (LSST) \citep{2012arXiv1211.0310L} which are listed in Sec.~4.4 {of F18}. As in the case for $C_\ell^{\phi\phi}$, we find that the uncertainty for $C_\ell^{\kappa g}$ in the SFB formalism decreases significantly at the same scales. In Table~\ref{tab:snr_kg}, we list the SNR values of the $\kappa g$ autocorrelation for the Plane-Parallel case and our SFB case. We find an increase of {only 10\%, mainly since the SNR is dominated by small scales where the PP noise prediction is already in parity with the signal $C_\ell^{\phi\phi}$ such that reducing the noise at large scales doesn't help much.  However, for lower redshift bins the increase is as high as 33\% and all the redshift bins should have higher increases when you just consider angular scales $\ell<100$ which may be relevant for certain cosmological applications.}
\begin{table}
    \centering
    \begin{tabular}{|c||c|c|c|}
        \cline{1-4}
          z & Plane-Parallel& SFB & \% Increase\\ \hline\hline
          $1.38\leq z\leq 1.48$ & 1.22  & 3.9  & 219  \\ \hline
          $1.48\leq z\leq 1.59$ & 1.27  & 3.2  & 151  \\ \hline
          $1.59\leq z\leq 1.71$ & 1.29  & 2.7  & 109  \\ \hline 
          $1.71\leq z\leq 1.84$ & 1.25  & 2.1  & 68  \\ \hline
          $1.84\leq z\leq 1.99$ & 1.25  & 2.0  & 60  \\ \hline
          $1.99\leq z\leq 2.16$ & 1.02  & 1.5  & 47 
          \\ \hline
          $2.16\leq z\leq 2.35$ & 0.88  & 1.3  & 47  \\ \hline
          $2.35\leq z\leq 2.57$ & 0.96  & 1.1  & 14  \\ \hline\hline
          $1.38\leq z\leq 2.57$ & 3.51  & 7.3  & 107 \\ \hline
    \end{tabular}
    \caption{Signal-to-noise ratio (SNR) of $C_\ell^{\phi\phi}$ for a HIRAX 21-cm intensity map at redshift 2 over a redshift bin size $\Delta z=0.16$ {over redshift range $1.38<z<2.57$} and LSST galaxies in the redshift range $0<z<5$.  Note that these SNR values are overestimated in that they do not account for bias-hardening to remove the effect of nonlinear clustering in the source maps {or the nonlinear clustering in the fiducial power spectrum, though we do limit the nonlinear scales used to $k_{\rm max}=0.5h$/Mpc}.  We also show the increase in the SNR when replacing the plane-parallel estimator with the SFB full-sky estimator, finding that the SNR changes significantly.}
    \label{tab:snr_pp}
\end{table}

\begin{table}
    \centering
    \begin{tabular}{|c||c|c|c|}
        \cline{1-4}
          z & Plane-Parallel& SFB& \% Increase\\ \hline\hline
          $1.38\leq z\leq 1.48$ & 23  & 31  & 33 \\ \hline
          $1.48\leq z\leq 1.59$ & 24  & 27  & 12.5 \\ \hline
          $1.59\leq z\leq 1.71$ & 25  & 26  & 3.8 \\\hline 
          $1.71\leq z\leq 1.84$ & 25  & 26  & 3.8 \\ \hline
          $1.84\leq z\leq 1.99$ & 26  & 26  & 3.8 \\\hline
          $1.99\leq z\leq 2.16$ & 24  & 25  & 4.16 \\\hline
          $2.16\leq z\leq 2.35$ & 23  & 25  & 8.6 \\\hline
          $2.35\leq z\leq 2.57$ & 25.1  & 25.3  & 0.7 \\ \hline\hline
          $1.38\leq z\leq 2.57$ & 69  & 76  & 10.1 \\ \hline
    \end{tabular}
    \caption{{Signal-to-noise ratio (SNR) of $C_\ell^{\kappa g}$ for a HIRAX 21-cm intensity map over redshift range $1.38<z<2.57$ and LSST galaxies in the redshift range $0<z<5$.  Note that the increase in the SNR for the SFB case is low overall, although the increase can be as high as 33\% for the lowest redshifts.}}
    \label{tab:snr_kg}
\end{table}

\begin{figure}
\begin{center}
\includegraphics[width=0.5\textwidth]{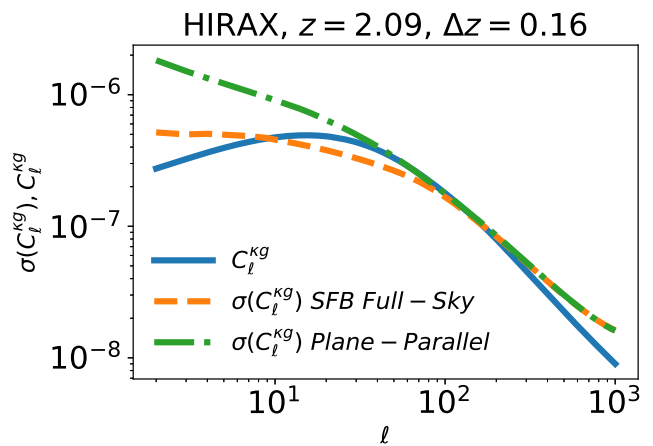}
\caption{\label{F:clkg}  The same as Fig.~\ref{F:clpp} except for $C_\ell^{\kappa g}$.  We find that the noise using the full-sky estimator decreases significantly at large scales.}
\end{center}
\end{figure}

We expect the difference in {these} estimators to affect the science yield for upcoming 21-cm lensing surveys. {However, this will be limited by wide-angle systematics in 21-cm intensity maps.  CMB lensing maps from the Planck satellite are currently limited to $\ell>8$ due to systematics, and we expect this issue to be worse for 21-cm intensity maps.  However, it may be possible for a futuristic survey with a significant increase in the number of detectors can achieve a high enough angular resolution to mitigate these systematic effects.  Even if this is not the case, it would only reduce the utility of the SFB full sky estimator since the SFB flat-sky predictions differ from PP at even fairly small angular scales $\ell\sim100$.}

One application that can benefit from a {full-sky estimator} is constraints on modified gravity through measurements of $E_G$ \citep{2015MNRAS.449.4326P} using 21-cm lensing \citep{2016MNRAS.461.1457P}.  In particular, modes {$\ell\lesssim 100$} could be relevant in these constraints and thus it would be beneficial to consider $E_G$ constraints with these new estimators.  Also, this estimator is not only applicable to 21-cm lensing, but any emission map lensing that can be mapped over large areas and multiple redshifts.  This estimator will not readily be useful for upcoming intensity maps of galactic emission lines such as CO, CII, or Ly$\alpha$ since they tend to map over small areas.  However, it may be beneficial for some science cases to sacrifice sensitivity to map a large area,  in which case a large-scale lensing survey for these lines may be feasible, particularly if a futuristic survey could map larger areas while maintaining current sensitivities.

Finally, while we only construct an estimator for a 2D lensing field $\phi_{\ell m}$, this work could be readily extended to {probe} $\phi_{\ell m}(k)$, the SFB transform of the 3D lensing field as a function of the radial distance to the source.  3D lensing in SFB coordinates have already been considered and measured for galaxy lensing surveys \citep{2005PhRvD..72b3516C,2014MNRAS.442.1326K}, and in principle this could also be applied to {LIM} lensing as well.  A major advantage of this approach would be that while {$C_\ell^{\kappa g}$, the angular lensing-convergence-galaxy cross-power spectrum, is a projection of the cross-power spectrum $P^{\nabla^2\Phi g}(k)$, the SFB cross-power spectrum $C_\ell^{\kappa g}(k,k')$ is directly proportional to $P^{\nabla^2\Phi g}(k)$.}  This would especially be advantageous for measurements of $E_G$, in that the $E_G$ estimator would no longer require line-of-sight integrals over the numerator and denominator, eliminating systematics associated with the redshift distribution as well as naturally probing scale-dependent growth.  We leave this for future work.

\section{Conclusions}\label{S:conc}

In this paper, we considered full sky effects on LIM lensing, particularly considering HIRAX, an upcoming survey for the 21 cm line. We have conducted our analysis using the spherical Fourier-Bessel series expansion of {three-dimensional} functions. Isotropy broken by weak lensing induces correlations which can be used to reconstruct the lensing potential by using a quadratic estimator. For large scales in particular, curved geometry becomes relevant through the extra correlations due to the three dimensional nature of the LIM signal, which is an additional benefit on top of the use of information at multiple redshifts.

In our forecasts of lensing reconstruction noise for the predicted 21 cm power spectra from HIRAX we include both our SFB Full-Sky and Flat-Sky estimators, in which we notice a significant drop in lensing noise for large scales {($\ell \lesssim 100$)}, scales past which we also notice a deviation between the {SFB and PP lensing} noise, which is suggestive of the key role of geometry in this effect. In addition, in our forecasts of $C_\ell^{\kappa g}$, where we assume galaxy measurements from the future LSST survey, we notice a similar trend at roughly the same angular scales. Conversely, at sufficiently small angular scales ($\ell \geqslant 100$), the difference between flat and full sky estimators is negligible, as expected. The corresponding SNR reflect this difference from the Plane-Parallel treatment of 3D weak lensing. Indeed, we find an increase in SNR by {107\%} for $C_\ell^{\phi\phi}$ and {10\%} for $C_\ell^{\kappa g}$, both in relation to the corresponding Plane-Parallel case.

{Considering our formula for the angular scale at which the SFB and PP predictions deviate, $L_{\rm dev}=\ell_{\rm res}-k_{\rm eq}R$, we expect this effect to be relavant mainly for low-redshift LIM surveys.  For high-redshift surveys, not only does $k_{\rm eq}R$ increase, but $\ell_{\rm res}$ decreases since the angular resolution is lower for larger observed wavelengths.  Both of these effects would lower $L_{\rm dev}$.  Another effect is that $L_{\rm dev}$ is dependent on cosmic parameters through both $k_{\rm eq}$ and $R(z)$.  Whether $L_{\rm dev}$ could be measured with enough precision to serve as a standard ruler we will leave as an exercise for the reader.}

Indeed, properly accounting for the spherical geometry shows promising benefits at the relevant scales. In particular, such a treatment is useful when one considers effects which dominate at large scales. Further, one may also consider full sky effects on reconstruction through polarization of the 21 cm line as well, which would be expected to demonstrate similar trends across large scales.

{Further}, we note that we have accounted for the fact that the survey is over a thin shell {by applying an approximate boundary condition at the minimum radius of the survey volume.  While this has worked reasonably well,} in order to properly do a spherical-shell calculation {of the lensing potential} one would have to use Bessel functions of both first and second type {as was done for the galaxy power spectrum in \citet{samushia},} which naturally account for the fewer modes in a spherical shell. {Finally, one should note that we have approximated $C_{\ell nn'}\approx C_{\ell n}\delta_{nn'}$ in our forecasts, which is reasonable for the non-RSD case. However, this is not strictly the case if one includes RSD (particularly for $k_{\ell n}R\sim\ell$). Properly factoring RSD effects would thus require using the full $C_{\ell nn'}$. Regardless, we expect that the overall trends of our results should be immune to this detail.} We leave a full investigation of these effects for future work.

\section*{Acknowledgements}

We wish to thank {S.~Foreman,} T.~Kitching, B.~Leidstedt, {A.~Liu,} and L.~Samushia for helpful comments and discussions.  {We also thank E.~Schaan for reviewing an earlier manuscript, and we also thank the referee for their comments.}  AP was supported by NASA ROSES NNH17ZDA001N-ATP Grant No. 80NSSC18K10148.  PC was supported by the Dean's Undergraduate Research Fund at New York University.

%%%%%%%%%%%%%%%%%%%%%%%%%%%%%%%%%%%%%%%%%%%%%%%%%%



\bibliographystyle{mnras}
%\bibliography{refs.bib} % if your bibtex file is called example.bib

%\begin{thebibliography}{99}
%\end{thebibliography}

 \newcommand{\noop}[1]{}

%%%%%%%%%%%%%%%%%%%%%%%%%%%%%%%%%%%%%%%%%%%%%%%%%%

%%%%%%%%%%%%%%%%% APPENDICES %%%%%%%%%%%%%%%%%%%%%

\appendix

\section{Quadratic Estimator for Lensing in the Flat-Sky Limit}
\label{app:A}
Here, we derive the quadratic estimator of the lensing potential in the flat-sky limit, where we require to move from the SFB moments of the 21 cm signal to its Fourier moments in the limit of small angles. 
We use the following approximation {from \cite{2003PhRvD..67h3002O}} for a function defined on a sphere in order to relate it with its harmonic moments on the sphere
\begin{eqnarray}
f(\ell)\approx\sqrt[]{\frac{4\pi}{2\ell+1}}\sum_{m}i^{-m}f_{\ell m}e^{im\varphi_{\ell}}\, .
\end{eqnarray}
We apply this dictionary to Eq.~\ref{eq:phihat} to approximate the flat-sky estimator for the lensing potential as
{\begin{eqnarray}
\hat{\phi}(\ell)&\approx &A(\ell)\sqrt[]{\frac{4\pi}{2\ell+1}}\sum_{m}i^{-m}e^{im\varphi_{\ell}}\sum_{\ell_1 m_1 n_1}\sum_{\ell_2 m_2 n_2}(-1)^mi^{m_1+m_2}\nonumber\\
&&\times\left( {\begin{array}{ccc}
   \ell_1 & \ell_2 & \ell\\
   m_1 & m_2 & -m\\ 
   \end{array}} \right)\frac{\sqrt[]{\ell_1 \ell_2}}{2\pi}\int_0^{2\pi}\int_0^{2\pi}\frac{d\varphi_{\ell_1}}{2\pi}\frac{d\varphi_{\ell_2}}{2\pi}\nonumber \\ 
   &&\times e^{-im_1\varphi_{\ell_1}}e^{-im_2\varphi_{\ell_2}}g_{\ell_1\ell_2\ell}^{n_1 n_2}\widetilde{T}_{n_1}( \boldsymbol{\ell_1})\widetilde{T}_{n_2}(\boldsymbol{\ell_2})\, .\nonumber \\
\label{eq:phiell1}
\end{eqnarray}}
We require an additional approximation in the limit $\ell,\ell_1,\ell_2\gg 1$ \citep{2003PhRvD..67h3002O}
\begin{eqnarray}
\ell(\ell+1)+\ell_1(\ell_1+1)-\ell_2(\ell_2+1)\approx 2\boldsymbol{\ell}\cdot\boldsymbol{\ell_1}\, .
\end{eqnarray}
Thus, we extract the geometric factors in $f_{\ell_1\ell\ell_2}^{n_1 n_2}$ and $g_{\ell_1\ell_2\ell}^{n_1 n_2}$ and make the following redefinitions in terms of corresponding barred quantities

\begin{eqnarray}
f_{\ell_1\ell\ell_2}^{n_1 n_2}&=&\sqrt[]{\frac{(2\ell_1+1)(2\ell+1)(2\ell_2+1)}{4\pi}}\left( {\begin{array}{ccc}
\ell & \ell_1 & \ell_2 \\
0 & 0 & 0 \\
\end{array}}\right)\bar{f}_{\ell_1\ell\ell_2}^{n_1 n_2}\nonumber \\
g_{\ell_1\ell_2\ell}^{n_1 n_2}&=&\sqrt[]{\frac{(2\ell_1+1)(2\ell+1)(2\ell_2+1)}{4\pi}}\left( {\begin{array}{ccc}
\ell & \ell_1 & \ell_2 \\
0 & 0 & 0 \\
\end{array}}\right)\bar{g}_{\ell_1\ell_2\ell}^{n_1 n_2}\, ,\nonumber \\
\end{eqnarray}
where
{\begin{eqnarray}
\bar{f}_{\ell_1\ell\ell_2}^{n_1 n_2}= M_{\ell_1 \ell_2}^{n_1 n_2} \boldsymbol{\ell}\cdot\boldsymbol{\ell_1}+M_{\ell_2 \ell_1}^{n_2 n_1} \boldsymbol{\ell}\cdot\boldsymbol{\ell_2}\, .
\end{eqnarray}}
We will specify the approximations for $\bar{g}_{\ell_1\ell_2\ell}^{n_1 n_2}$ later since they require additional machinery. \newline 
{We use the formula for the integral of three spherical harmonics,}
{\begin{eqnarray}\label{E:3ylm}
\int&d\Omega& Y_{\ell_1m_1}(\hatr)Y_{\ell_2m_2}(\hatr)Y_{\ell_3m_3}(\hatr)\nonumber\\
&=&\sqrt{\frac{(2\ell_1+1)(2\ell_2+1)(2\ell_3+1)}{4\pi}}\left( {\begin{array}{ccc}
\ell & \ell_1 & \ell_2 \\
0 & 0 & 0 \\
\end{array}}\right)\nonumber\\
&&\times\left( {\begin{array}{ccc}
\ell & \ell_1 & \ell_2 \\
m_1 & m_2 & m_3 \\
\end{array}}\right)\, ,
\end{eqnarray}}
and make use of the approximation \citep{2003PhRvD..67h3002O}

\begin{eqnarray}
e^{i\boldsymbol{\ell}\cdot\boldsymbol{\hat{n}}}\approx\sqrt[]{\frac{2\pi}{\ell}}\sum_{m}i^{m}Y_{\ell m}e^{-i\varphi_{\ell}m}\, .
\label{eq:planewaveflat}
\end{eqnarray}
This allows us to simplify Eq (\ref{eq:phiell1}) to
{\begin{eqnarray}
\hat{\phi}(\boldsymbol{\ell})&\approx & A(\ell)\sum_{n_1 n_2}\int\frac{d^2\ell_1}{(2\pi)^2}\int d^2\ell_2\,\bar{g}_{\ell_1\ell_2\ell}^{n_1 n_2}\nonumber \\ 
   &&\times\widetilde{T}_{n_1}(\boldsymbol{\ell}_1)\widetilde{T}_{n_2}(\boldsymbol{\ell}_2)\delta_D^2(\vec{\ell}_2+\vec{\ell}_1-\vec{\ell})\nonumber\\
   &=&A(\ell)\sum_{n_1 n_2}\int\frac{d^2\ell_1}{(2\pi)^2}\bar{g}_{\ell_1\ell_2\ell}^{n_1 n_2}\widetilde{T}_{n_1}(\boldsymbol{\ell}_1)\nonumber\\
   &&\times\widetilde{T}_{n_2}(\boldsymbol{\ell}_2)\, ,
\end{eqnarray}}
{where $\boldsymbol{\ell_2}\equiv\boldsymbol{\ell}-\boldsymbol{\ell_1}$ and the same for the primed case.}  Following the same procedure as for the full-sky case, we take the covariance of the lensing estimator, which is given as
{\begin{eqnarray}
\langle\hat{\phi}(\boldsymbol{\ell})\hat{\phi}^*(\boldsymbol{\ell}')\rangle &=&A(\ell)A(\ell')\sum_{n_1 n_2}\sum_{n_1' n_2'}\int\frac{d^2\ell_1}{(2\pi)^2}\int\frac{d^2\ell_1'}{(2\pi)^2}\nonumber \\
&&\times\bar{g}_{\ell_1\ell_2\ell}^{n_1 n_2}\bar{g}_{\ell_1'\ell_2'\ell'}^{n_1' n_2'}\langle\widetilde{T}_{n_1}(\boldsymbol{\ell}_1)\widetilde{T}_{n_2}(\boldsymbol{\ell}_2)\nonumber \\ 
   &&\times\widetilde{T}_{n_1'}(\boldsymbol{\ell}_1')\widetilde{T}_{n_2'}(\boldsymbol{\ell}_2')\rangle\nonumber \\
&=&(2\pi)^2\delta_D^2(\boldsymbol{\ell}-\boldsymbol{\ell}')A(\ell)A(\ell')\sum_{n_1 n_2}\sum_{n_1' n_2'}\int\frac{d^2\ell_1}{(2\pi)^2}\nonumber \\
&&\times\bar{g}_{\ell_1\ell_2\ell}^{n_1 n_2}(\bar{g}_{\ell_1\ell_2\ell}^{n_1' n_2'}C_{\ell_1 n_1n_1'}^{\text{tot}}C_{\ell_2n_2n_2'}^{\text{tot}} \nonumber\\
&&+\bar{g}_{\ell_2\ell_1\ell}^{n_1' n_2'}C_{\ell_1 n_1n_2'}^{\text{tot}}C_{\ell_2n_2n_1'}^{\text{tot}})\, .
\end{eqnarray}}
%{\begin{eqnarray}
%\langle\hat{\phi}(\vec{\ell})\hat{\phi}^*(\vec{\ell}')\rangle &=&A(\ell)A(\ell')\sum_{n_1 n_2}\sum_{n_1' n_2'}\int\frac{d\vec{\ell_1}}{(2\pi)^2}\int\frac{d\vec{\ell_1}'}{(2\pi)^2}\nonumber \\
%&&\times\bar{g}_{\ell_1\ell_2\ell}^{n_1 n_2}(\ell_1,\ell-\ell_1)\bar{g}^{n_1' n_2'}(\ell_1',\ell'-\ell_1')\nonumber \\
%&&\times \langle\widetilde{T}(k_{\ell_1 n_1},\ell_1)\widetilde{T}(k_{\ell-\ell_1 \, n_2},\ell-\ell_1)\nonumber \\ 
%   &&\times\widetilde{T}(k_{\ell_1' n_1'},\ell_1')\widetilde{T}(k_{\ell'-\ell_1' \, n_2'},\ell'-\ell_1')\rangle\nonumber \\
%&=&(2\pi)^2\delta_D^2(\vec{\ell}-\vec{\ell}')A(\ell)A(\ell')\sum_{n_1 n_2}\int\frac{d\vec{\ell_1}}{(2\pi)^2}\nonumber \\
%&&\times\bar{g}^{n_1 n_2}(\ell_1,\ell-\ell_1) C^{\text{tot}}(k_{\ell \,n_1},\ell_1)\nonumber\\
%&&\times C^{\text{tot}}(k_{|\ell-\ell_1| \,n_2},\ell-\ell_1)(\bar{g}^{n_1 n_2}(\ell_1,\ell-\ell_1)\nonumber\\
%&&+\bar{g}^{n_2 n_1}(\ell-\ell_1,\ell_1))\nonumber \\
%\end{eqnarray}}
%Here, $C_{\ell nn'}^{\text{tot}}$ is the flat-sky power spectrum and it is defined as
%\begin{eqnarray}
%C^{\text{tot}}(k_{\ell n},\ell)\equiv C(k_{\ell n},\ell)+N(k_{\ell n},\ell)\, .
%\end{eqnarray}
Thus, the variance is

{\begin{eqnarray}
\langle|\hat{\phi}(\boldsymbol{\ell})|^2\rangle &=& (2\pi)^2\delta_D^2(0)A^2(\ell)\sum_{n_1 n_2}\sum_{n_1'n_2'}\int\frac{d^2\ell_1}{(2\pi)^2}\bar{g}_{\ell_1\ell_2\ell}^{n_1 n_2}\bar{g}_{\ell_1\ell_2\ell}^{n_1' n_2'}\nonumber\\
&&\times 2C_{\ell_1 n_1n_1'}^{\text{tot}}C_{\ell_2 n_2n_2'}^{\text{tot}}\, .
\end{eqnarray}}
We are required to approximate $A(\ell)$ as well, which we do below. We expand Eq.~\ref{eq:Aell} to obtain

\begin{eqnarray}
A_\ell^{-1}&=&\frac{1}{2\ell+1}\sum_{\ell_1\ell_2}\sum_{n_1 n_2}\frac{(2\ell+1)(2\ell_1+1)(2\ell_2+1)}{4\pi}\nonumber\\
&&\times\left( {\begin{array}{ccc}
   \ell_1 & \ell & \ell_2\\
   0 & 0 & 0\\ 
   \end{array}} \right)^2\bar{g}^{n_1 n_2}_{\ell_1\ell_2\ell}\bar{f}^{n_1n_2}_{\ell_1\ell\ell_2}\nonumber\\
   &\approx &\sum_{n_1n_2}\sum_{\ell_1\ell_2}\frac{\ell_1\ell_2}{\pi}\left( {\begin{array}{ccc}
   \ell_1 & \ell & \ell_2\\
   0 & 0 & 0\\ 
   \end{array}} \right)^2\bar{g}^{n_1 n_2}_{\ell_1\ell_2\ell}\bar{f}^{n_1n_2}_{\ell_1\ell\ell_2}\, . \nonumber \\
\end{eqnarray}
{We use the formula for the integral of three spherical harmonics in Eq.~\ref{E:3ylm},} which in the limit $\ell,\ell_1,\ell_2\gg 1$ produces

\begin{multline}
\frac{1}{4\pi}\int d\hat{\textbf{r}}\left(\sqrt{\frac{2\pi}{\ell}}Y^*_{\ell 0}(\hat{\textbf{r}})\right)\left(\sqrt{\frac{2\pi}{\ell_1}}Y_{\ell_1 0}(\hat{\textbf{r}})\right)\left(\sqrt{\frac{2\pi}{\ell_2}}Y_{\ell_2 0}(\hat{\textbf{r}})\right) \\
\approx\left( {\begin{array}{ccc}
   \ell_1 & \ell & \ell_2\\
   0 & 0 & 0\\ 
   \end{array}} \right)^2\, .
\end{multline}
Further, note that
{\begin{eqnarray}
\sqrt{\frac{2\pi}{\ell}}Y^*_{\ell 0}(\hat{\textbf{r}})&=&\sqrt{\frac{2\pi}{\ell}}\sum_{m}i^{-m}Y^*_{\ell m}(\hat{\textbf{r}})\delta_{m,0}\nonumber\\
&\approx &\sqrt{\frac{2\pi}{\ell}}\sum_{m}i^{-m}Y^*_{\ell m}(\hat{\textbf{r}})\int \frac{d\varphi_{\ell}}{2\pi}e^{im\varphi_\ell}\nonumber\\
&=&\int \frac{d\varphi_{\ell}}{2\pi}\sqrt{\frac{2\pi}{\ell}}\sum_{m}i^{-m}Y^*_{\ell m}(\hat{\textbf{r}})e^{im\varphi_\ell}\nonumber\\
&\approx &\int \frac{d\varphi_{\ell}}{2\pi}e^{i\boldsymbol{\ell}\cdot\boldsymbol{\hat{n}}} \, ,
\end{eqnarray}}
where we have used Eq.~\ref{eq:planewaveflat}. Using this, we can obtain the result for $A(\ell)$

{\begin{eqnarray}
A(\ell)^{-1}&=&\sum_{n_1n_2}\int \frac{d^2\ell_1}{(2\pi)^2}\int d^2\ell_2\bar{g}^{n_1 n_2}_{\ell_1\ell_2\ell}\bar{f}^{n_1 n_2}_{\ell_1\ell\ell_2}\nonumber \\
&&\times \delta_D^2(\boldsymbol{\ell}_1+\boldsymbol{\ell}_2-\boldsymbol{\ell})\nonumber\\
&=&\sum_{n_1 n_2}\int\frac{d^2\ell_1}{(2\pi)^2}\bar{g}^{n_1 n_2}_{\ell_1\ell_2\ell}\bar{f}^{n_1 n_2}_{\ell_1\ell\ell_2}\, . \nonumber \\
\end{eqnarray}}
Similarly, we arrive at the flat-sky expression for {$\bar{g}^{n_1 n_2}_{\ell_1\ell_2\ell}$}

{\begin{eqnarray}
\bar{g}^{n_1 n_2}_{\ell_1\ell_2\ell}=\frac{1}{2}\sum_{n_1'n_2'}(C_{\ell_1}^{\rm tot}{}^{-1})_{n_1n_1'}(C_{\ell_2}^{\rm tot}{}^{-1})_{n_2n_2'}\bar{f}_{\ell_1\ell \ell_2}^{n_1'n_2'}\, .
\end{eqnarray}}
Finally, assuming a fiducial survey allows us to obtain the noise power spectrum

{\begin{eqnarray}
N^{\phi\phi}(\ell)^{-1}&=&\frac{1}{2}\sum_{n_1 n_2}\sum_{n_1'n_2'}\int\frac{d^2\ell_1}{(2\pi)^2}(C_{\ell_1}^{\rm tot}{}^{-1})_{n_1n_1'}(C_{\ell_2}^{\rm tot}{}^{-1})_{n_2n_2'}\nonumber\\
&&\times \bar{f}_{\ell_1\ell\ell_2}^{n_1'n_2'}\bar{f}_{\ell_1\ell\ell_2}^{n_1n_2}
\end{eqnarray}}
%where,
%\begin{equation}
%\bar{f}_{\ell_1\ell\ell_2}^{n_1 n_2}=\tau_{\ell_1 n_1}C(k_{\ell_1 n_1},\ell_1) \boldsymbol{\ell}\cdot\boldsymbol{\ell_1}+\tau_{\ell_2, n_2}C(k_{\ell_2 \, n_2},\ell_2) \boldsymbol{\ell}\cdot\boldsymbol{\ell_2}
%\end{equation}

\section{Intensity Mapping Instrumental Effects}\label{S:intinst}

\subsection{Angular and Radial Window Functions}

Just like for the CMB, a line intensity map signal will be convolved by the instrument.  Yet while a CMB map is only convolved by the beam in two dimensions, an intensity map will also be convolved along the line-of-sight due to the spectral window in frequency space.  In this section, we derive the angular (beam) and radial (spectral) window functions that can be applied to the {LIM} signal. {Throughout this derivation, any integrals over $r$ will be understood to be over the range $r_{\rm min}<r<r_{\rm max}$.}

We start by defining a window function in real space $D(\vecr'|\vecr)$ such that the observed map $T^{\rm obs}(\vecr)$ is a convolution of the true map $T(\vecr)$ with the window function, assuming no instrumental noise or masking, according to
\begin{eqnarray}
T^{\rm obs}(\vecr)=\int d^3r'\,D(\vecr'|\vecr)T(\vecr')\, .
\end{eqnarray}
Next, we assume that the angular and radial window functions are separable such that $D(\vecr'|\vecr)=A(r'|r)B(\hatr'|\hatr)$ where $A(r'|r)$ and $B(\hatr'|\hatr)$ are the radial and angular window functions, respectively.  This allows us to rewrite the convolution as
\begin{eqnarray}
T^{\rm obs}(\vecr)=\int dr'r'^2\,A(r'|r)\int d\Omega B(\hatr'|\hatr)T(\vecr')\, .
\end{eqnarray}
It can be shown that the SFB moments of the observed map can be written in the form
\begin{eqnarray}
T_{\ell m n}^{\rm obs}&=&\sum_{\ell'm'n'}\tau_{\ell'n'}T_{\ell'm'n'}\int dr\,r^2\,dr'\,r'^2j_\ell(k_{\ell n}r)A(r'|r)j_{\ell'}(k_{\ell'n'}r')\nonumber\\
&&\times\int d\Omega\,d\Omega'\,Y_{\ell m}^*(\hatr)B(\hatr'|\hatr)Y_{\ell'm'}(\hatr')\, .
\end{eqnarray}

By taking the covariance of these moments, we find
{
\begin{widetext}
\begin{eqnarray}\label{E:tlmntlmnwin}
\VEV{T_{\ell_1 m_1 n_1}^{\rm obs}T_{\ell_2 m_2 n_2}^{\rm obs,*}}&=&\sum_{\ell'm'n_1'n_2'}\tau_{\ell'n_1'}\tau_{\ell'n_2'}C_{\ell'n_1'n_2'}\int dr_1\,r_1^2\,dr_2\,r_2^2 j_{\ell'}(k_{\ell'n_1'}r_1)j_{\ell'}(k_{\ell'n_2'}r_2)j_{\ell_1}(k_{\ell_1n_1}r_1)j_{\ell_2}(k_{\ell_2n_2}r_2)\nonumber \\
&&\times\int d\Omega_1\,d\Omega_2\,Y_{\ell_1 m_1}^*(\hatr_1)Y_{\ell_2m_2}(\hatr_2)Y_{\ell' m'}^*(\hatr_1) Y_{\ell'm'}(\hatr_2) W_{\ell'n_1'n_2'}^A(r_1,r_2)W^B_{\ell'}(\hatr_1,\hatr_2)\, ,
\end{eqnarray}
\end{widetext}
}
where
\begin{eqnarray}
W_{\ell nn'}^A(r_1,r_2)&=&\left(\frac{\int dr\,r^2j_\ell(k_{\ell n_1}r)A(r|r_1)}{j_\ell(k_{\ell n_1}r_1)}\right)\nonumber\\
&&\times\left(\frac{\int dr\,r^2j_\ell(k_{\ell n_2}r)A(r|r_2)}{j_\ell(k_{\ell n_2}r_2)}\right)\, ,
\end{eqnarray}
and
\begin{eqnarray}
W_{\ell}^B(r_1,r_2)=\frac{1}{P_\ell(\hatr_1\cdot\hatr_2)}\int d\Omega\,d\Omega'P(\hatr\cdot\hatr')B(\hatr|\hatr_1)B(\hatr'|\hatr_2)\, .
\end{eqnarray}
A simpler expression for the mode covariance arises if it is assumed that $A(r'|r)$ and $B(\hatr'|\hatr)$ are both highly peaked, in which case $W_{\ell nn'}^A$ and $W_\ell^B$ should be independent of pixel locations $\vecr_1$ and $\vecr_2$.  In this case, the integrals in Eq.~\ref{E:tlmntlmnwin} can be performed, giving us
\begin{eqnarray}
\VEV{T_{\ell_1 m_1 n_1}^{\rm obs}T_{\ell_2 m_2 n_2}^{\rm obs,*}}=W_{\ell_1n_1n_2}^AW_{\ell_1}^BC_{\ell_1n_1n_2}\delta_{\ell_1\ell_2}\delta_{m_1m_2}
\end{eqnarray}

For the analytical formulae, this representation for the radial and angular harmonic window functions will be retained.  In Appendix \ref{S:specwin} we derive that the radial harmonic window function can be written as a sinc function in the form
\begin{eqnarray}\label{E:wna_app}
W_{\ell nn'}^A=A(k_{\ell n})A(k_{\ell n'})\, ,
\end{eqnarray}
where
\begin{eqnarray}\label{E:acalc_app}
A(k)=\frac{\pi}{4}\left[{\rm sinc}\left(\frac{k\Delta r-\pi}{2}\right)+{\rm sinc}\left(\frac{k\Delta r+\pi}{2}\right)\right]\, .
\end{eqnarray}
where $\Delta r$ is the full width of the radial window.

%For the forecasts, we will assume a Gaussian angular beam, such that in the small-angle limit, we can write $W_\ell^B=e^{\ell^2\sigma_b^2}$, where $\sigma_b$ is the angular resolution just like for the CMB.  Along the radial direction, we will assume a convenient spectral window function given by
%\begin{eqnarray}\label{E:ar}
%A(r'|r)=\left\{\begin{array}{ll}\frac{\pi}{2\Delta r\,r^2}\cos\left[\frac{\pi(r'-r)}{\Delta r}\right]&r-\Delta r/2<r'<r+\Delta r/2\\0&{\rm otherwise}\end{array}\right.\, ,
%\end{eqnarray}
%where $\Delta r$ is the full width of the radial window.  This is similar to the window assumed in L17, except we add an extra pre-factor to make it normalize conveniently.  

\subsection{Spectral Window Function}\label{S:specwin}

Here we derive the radial harmonic window function $W_{\ell nn'}^A=A_\ell(k_{\ell n})A_\ell(k_{\ell n'})$, where
\begin{eqnarray}\label{E:alk}
A_\ell(k)=\frac{\int dr' r'^2A(r'|r)j_\ell(kr')}{j_\ell(kr)}\, ,
\end{eqnarray}
and $A(r'|r)$ is given by
{\begin{eqnarray}\label{E:ar}
A(r'|r)=\left\{\begin{array}{ll}\frac{\pi}{2\Delta r\,r'^2}\cos\left[\frac{\pi(r'-r)}{\Delta r}\right]&r-\Delta r/2<r'<r+\Delta r/2\\0&{\rm otherwise}\end{array}\right.\, ,
\end{eqnarray}
where $\Delta r$ is the full width of the radial window.  This is similar to the window assumed in L17, except we add an extra pre-factor to make it normalize conveniently.  Technically the integration over $r$ should be over the range $r_{\rm min}<r<r_{\rm max}$; however, since the window function A(r'|r) is expected to be relatively narrow, we can extend the integration range to $r>0$ without losing accuracy.} We are mainly going to consider the case where {the modes are more radial than angular}, $\ell\ll kr$.  However, for the case where $\ell\gtrsim kr$,  $j_\ell(kr)$ is a smooth, monotonic function that can be taken out of the integral compared to the highly-peaked $A(r'|r)$, which itself is normalized.  Thus, for $\ell\gtrsim kr$, $A_\ell(k)\simeq1$.

For $\ell\ll kr$, we can use the approximation
\begin{eqnarray}
j_\ell(x)=\frac{1}{x}\sin\left(x-\frac{\pi\ell}{2}\right)\, ,
\end{eqnarray}
allowing us to write the numerator of Eq.~\ref{E:alk}, which we label $Q$, as
\begin{eqnarray}
Q=\frac{1}{k}\int_0^\infty dr'\,r'\sin\left(kr'-\frac{\pi\ell}{2}\right)A(r'|r)\, .
\end{eqnarray}
Since $A(r'|r)$ is cut off for $r'>r+\Delta r/2$, it is fine to extend the integral out to infinity.  We evaluate this integral using a similar trick as was performed in L17.  First, we convert the sine function to an exponential with a derivative.
\begin{eqnarray}
Q=-\frac{1}{k^2}\frac{\partial}{\partial\alpha}\left\{{\rm Re}\left[\int_0^\infty dr'\,e^{-i\alpha kr'+i\pi\ell/2}A(r'|r)\right]\right\}_{\alpha=1}\, .
\end{eqnarray}
Then, we define a new function $C(r')=A(r'+r|r)$ which can be written as
\begin{eqnarray}
C(r'|r)\left\{\begin{array}{ll}\frac{\pi}{2\Delta r\,(r+r')^2}\cos\left[\frac{\pi r}{\Delta r}\right]&-\Delta r/2<r'<+\Delta r/2\\0&{\rm otherwise}\end{array}\right.
\end{eqnarray}
Given this, we find
\begin{eqnarray}
Q&=&-\frac{1}{k^2}\frac{\partial}{\partial\alpha}\left\{{\rm Re}\left[\int_{-\infty}^\infty dx\,e^{-i\alpha kx}e^{-i\alpha kr+i\pi\ell/2}C(x)\right]\right\}_{\alpha=1}\nonumber\\
&=&-\frac{1}{k^2}\frac{\partial}{\partial\alpha}\left\{{\rm Re}\left[e^{-i\alpha kr+i\pi\ell/2}\tilde{C}(\alpha k)\right]\right\}_{\alpha=1}\, ,
\end{eqnarray}
where we define $x\equiv r'-r$ and we can extend the integral out to $-\infty$ since $C(x<-\Delta r/2)=0$, allowing to then define the Fourier transform $\tilde{C}(k)$.  Next, we make the argument that $C(x)$ is approximately an even function in the limit where $r'\ll r$.  Under this approximation, we can apply the Re operator to just the exponential, giving us
\begin{eqnarray}
Q&=&-\frac{1}{k^2}\frac{\partial}{\partial\alpha}\left[\cos\left(\alpha kr-\frac{\pi\ell}{2}\right)\tilde{C}(\alpha k)\right]_{\alpha=1}\nonumber\\
&=&\frac{1}{k}\left[r\sin\left(kr-\frac{\pi\ell}{2}\right)\tilde{C}(k)-\cos\left(kr-\frac{\pi\ell}{2}\right)\frac{\partial\tilde{C}}{\partial k}\right]\, .
\end{eqnarray}
Inserting this into Eq.~\ref{E:alk}, we can write
\begin{eqnarray}
A_\ell(k)=r^2\tilde{C}(k)-r\cot\left(kr-\frac{\pi\ell}{2}\right)\frac{\partial\tilde{C}}{\partial k}\, .
\end{eqnarray}

Now we evaluate $\tilde{C}(k)$.  Using the $r'<<r$ approximation, which allows us to just evaluate the real part of the Fourier transform, we find
\begin{eqnarray}
\tilde{C}(k)&=&\frac{\pi}{2\Delta r\,r^2}\int_{-\Delta r/2}^{\Delta r/2}dr\cos(kr)\cos\left(\frac{\pi r}{\Delta r}\right)\nonumber\\
&=&\frac{\pi}{4r^2}\left[{\rm sinc}\left(\frac{k\Delta r-\pi}{2}\right)+{\rm sinc}\left(\frac{k\Delta r+\pi}{2}\right)\right]\, .
\end{eqnarray}
From this, we know $\partial\tilde{C}/\partial k\propto \Delta r/r\ll1$ such that we can approximate $A_\ell(k)\simeq r^2\tilde{C}(k)$, which gives us Eq.~\ref{E:acalc_app}.

\subsection{Instrumental Noise}

Here we derive the instrumental noise contribution to the covariance of the SFB moments.  In this analysis we will set each three-dimensional pixel, or voxel, to be labeled by an index such that the noise fluctuations are labeled $T_i^N$.  We assume that the noise fluctuations are uncorrelated between voxels and that their distribution varies only with frequency, and thus the radial coordinate for a given line.  This gives us a noise covariance matrix $\VEV{T_i^NT_j^N}=\sigma_{N,i}^2\delta_{ij}$.

We begin by writing the SFB moment for the noise fluctuations as a sum over voxels
\begin{eqnarray}
T_{\ell m n}^N=\sum_i\Delta_i Y_{\ell m}^*(\hatr_i)j_\ell(k_{\ell n}r_i)T_i^N\, ,
\end{eqnarray}
where $\Delta_i$ is the volume of voxel $i$, given by $\Delta_i=\Delta\Omega\Delta r_ir_i^2$.  Taking the covariance, we find
\begin{eqnarray}
\VEV{T_{\ell m n}^NT_{\ell' m' n'}^N}&=&\sum_i\Delta_i^2 Y_{\ell m}^*(\hatr_i)Y_{\ell' m'}(\hatr_i)\nonumber\\
&&\times j_\ell(k_{\ell n}r_i)j_{\ell'}(k_{\ell' n'}r_i)\sigma_{N,i}^2\, .
\end{eqnarray}
Converting back from a sum to an integral, we can write the covariance, after integrating over angle, as
\begin{eqnarray}
\VEV{T_{\ell m n}^NT_{\ell' m' n'}^N}=N_{\ell nn'}\delta_{\ell\ell'}\delta_{mm'}\, ,
\end{eqnarray}
where
\begin{eqnarray}
N_{\ell nn'}=\int_{r_{\rm min}}^{r_{\rm max}} dr\,r^2\Delta(r)j_\ell(k_{\ell n}r)j_{\ell}(k_{\ell n'}r)\sigma_N^2(r)\, ,
\end{eqnarray}
and $\Delta(r)$ is the size of the voxel as a function or $r$.  The voxel depth is determined by the frequency bin size, such that we can write
\begin{eqnarray}
\Delta(r)=\Delta\Omega\Delta\nu\left|\frac{dr}{d\nu}\right|r^2\, ,
\end{eqnarray}
where
\begin{eqnarray}
\left|\frac{dr}{d\nu}\right|=\frac{c[1+z(r)]}{\nu H[z(r)]}\, .
\end{eqnarray}
Using the results from the previous section, we can now write
\begin{eqnarray}
\VEV{T_{\ell_1 m_1 n_1}^{\rm obs}T_{\ell_2 m_2 n_2}^{\rm obs,*}}=\left[W_{\ell_1 n_1n_2}^{AB}C_{\ell_1n_1n_2}+N_{\ell_1n_1n_2}\right]\delta_{\ell_1\ell_2}\delta_{m_1m_2}\, ,
\end{eqnarray}
where we define $W_{\ell n_1n_2}^{AB}\equiv W_{\ell n_1n_2}^{A}W_\ell^{B}$.

\section{Proof of Form of Quadratic Estimator}\label{A:estimator}

{Here we prove that our lensing estimator is the most general quadratic estimator that can be constructed for this problem.  Consider a general quadratic estimator for $\phi(\hat{n})$
\begin{eqnarray}
\hat{\phi}(\hat{n})&=&\int d^3x d^3y \widetilde{T}(\vec{x})\widetilde{T}(\vec{y})G(\vec{x},\vec{y},\hat{n}).
\end{eqnarray}}

{We may expand $\hat{\phi}$ and $\widetilde{T}$ in spherical harmonics and SFB series respectively to obtain
\begin{eqnarray}\label{E:est}
\hat{\phi}_{LM}&=&\sum_{\ell\ell' }\sum_{mm'}\sum_{nn'}\widetilde{T}_{\ell m n}\widetilde{T}_{\ell' m' n'}\tau_{\ell n}\tau_{\ell' n'}\nonumber \\
&&\times \int d^3x d^3y d\hat{n}j_{\ell}(k_{\ell n}x)j_{\ell'}(k_{\ell' n'}y) \nonumber \\
&&\times Y^*_{LM}(\hat{n})Y_{\ell m}(\hat{x})Y_{\ell' m'}(\hat{y})G(\vec{x},\vec{y},\hat{n}).
\end{eqnarray}}

{Now, $G(\vec{x},\vec{y},\hat{n})$ depends on three directions, thus it may be expanded in tripolar spherical harmonics as
\begin{eqnarray}
G(\vec{x},\vec{y},\hat{n})&=&\sum_{\ell_1\ell_2 L'}\sum_{m_1 m_2 M'}g_{\ell_1\ell_2L'}^{mm' M}(x,y)\left( {\begin{array}{ccc}
   \ell_1 & \ell_2 & L'\\
   m_1 & m_2 & M'\\ 
   \end{array}} \right)\nonumber \\
   &&\times Y_{\ell_1 m_1}(\hat{x})Y_{\ell_2 m_2}(\hat{y})Y_{L' M'}(\hat{n}).
\end{eqnarray}}

{One may integrate the angular part of Eq.~\ref{E:est} by using the orthogonality relations of spherical harmonics to obtain
\begin{eqnarray}
\hat{\phi}_{LM}&=&\sum_{\ell\ell' }\sum_{mm'}\sum_{nn'}(-1)^{\ell+\ell'+L+M}\tau_{\ell n}\tau_{\ell' n'}\widetilde{T}_{\ell m n}\widetilde{T}_{\ell' m' n'} \nonumber \\
&&\times\left( {\begin{array}{ccc}
   \ell & \ell' & L\\
   m & m' & -M\\ 
   \end{array}} \right)\int dx dy x^2 y^2j_{\ell}(k_{\ell n}x)j_{\ell'}(k_{\ell' n'}y)\nonumber\\
   &&\times g_{\ell \ell' L}^{mm'M}(x,y).
\end{eqnarray} 
Now, one may make the re-definition,
\begin{eqnarray}
g_{\ell \ell' L}^{mm'M, nn'}&\equiv&(-1)^{\ell+\ell'+L}\tau_{\ell n}\tau_{\ell' n'} \int dx dy x^2 y^2j_{\ell}(k_{\ell n}x)\nonumber\\
&&\times j_{\ell'}(k_{\ell' n'}y)g_{\ell \ell' L}(x,y)\, ,
\end{eqnarray}
which leaves the estimator in the form
\begin{eqnarray}
\hat{\phi}_{LM}&=&\sum_{\ell m n}\sum_{\ell' m' n'}(-1)^{M}\left( {\begin{array}{ccc}
   \ell & \ell' & L\\
   m & m' & -M\\ 
   \end{array}} \right) \nonumber\\ 
   &&\times g_{\ell\ell'L}^{mm'M,n n'}\widetilde{T}_{\ell m n}\widetilde{T}_{\ell' m' n'}.
\end{eqnarray}}

{But, $g_{\ell\ell'L}^{mm'M,nn'}$ is to be determined by minimizing $\langle\hat{\phi}_{LM}\hat{\phi}^*_{L'M'}\rangle$ which we assume to be diagonal, i.e., $\propto \delta_{LL'}\delta_{MM'}$. This can only be achieved due to the orthogonality of the 3j-symbol, so $g$ cannot depend on $\{m,m',M\}$. So we now have
\begin{eqnarray}
\hat{\phi}_{LM}&=&\sum_{\ell m n}\sum_{\ell' m' n'}(-1)^{M}\left( {\begin{array}{ccc}
   \ell & \ell' & L\\
   m & m' & -M\\ 
   \end{array}} \right) \nonumber\\ 
   &&\times g_{\ell\ell'L}^{n n'}\widetilde{T}_{\ell m n}\widetilde{T}_{\ell' m' n'}.
\end{eqnarray}
Finally, since we need an unbiased estimator, we need a suitable normalization which we denote as $A_L$. Thus we have the final form of the estimator as given in the main text.}

\section{Approximation for Spherical Bessel Function at large $\ell$}\label{A:saddle}
{Recall an integral representation of $j_{\ell}(x)$
\begin{eqnarray}
j_{\ell}(x)&=&\frac{1}{2}\frac{(x/2)^\ell}{\ell!}\int_{-1}^{1}dt\,e^{ixt}(1-t^2)^{\ell}.
\end{eqnarray}
It is known that performing a saddle-point approximation (along with a suitable substitution) produces the following approximation for $j_{\ell}(x)$ for the region $x\leq \ell$ ($\ell\gg 1$)
\begin{eqnarray}
j_{\ell}(x)&=&\frac{e^{\sqrt{\ell^2-x^2}}}{2x}\left(\frac{\ell-\sqrt{\ell^2-x^2}}{x}\right)^\ell\left(\frac{\ell-\sqrt{\ell^2-x^2}}{\sqrt{\ell^2-x^2}}\right)^{1/2}.
\end{eqnarray}
Similarly, one may use an integral representation for the spherical Hankel function $h_{\ell}^{(1)}(x)$ to saddle-point approximate it in the region $x\geq\ell$ ($\ell\gg1$)
\begin{eqnarray}
h^{(1)}_{\ell}(x)&=&\frac{e^{i\sqrt{x^2-\ell^2}}}{x}\left(\frac{\ell-i\sqrt{x^2-\ell^2}}{x}\right)^\ell\left(\frac{\ell-i\sqrt{x^2-\ell^2}}{i\sqrt{x^2-\ell^2}}\right)^{1/2}\, ,
\end{eqnarray}
where the real part of this expression gives the approximation for $j_\ell(x)$.  We may consider the regime $kr\geq \ell$ in particular with $x\equiv kr$. Firstly, note,
\begin{eqnarray}
\left(\frac{\ell-i\sqrt{x^2-\ell^2}}{i\sqrt{x^2-\ell^2}}\right)^{1/2}&=&i\left(1+i\frac{\ell}{2x}+\frac{\ell^3}{4x^3}+O(\ell^5/x^5))\right)\nonumber \\
&\approx& i.
\end{eqnarray}
Now we may define $k_{||}^2=k^2-\ell^2/r_1^2=k^2-k_{\perp}^2$. So we have
\begin{eqnarray}
\sqrt{x^2-\ell^2}&=&\sqrt{k_{||}^2r^2-\ell^2(1-r^2/r_1^2)} \nonumber \\
&\geq& \sqrt{k_{||}^2r^2-\ell^2(1-a^2)} \nonumber \\
&\approx& k_{||}r\, ,
\end{eqnarray}since $a^2=1+\frac{1}{N^2}$ where $N=\frac{R}{\Delta R}>>1$. In light of this, we have the following,
\begin{eqnarray}
e^{i\sqrt{x^2-\ell^2}}&\approx&e^{ik_{||}r}\, ,
\end{eqnarray}
and
\begin{eqnarray}
\left(\frac{\ell-i\sqrt{x^2-\ell^2}}{x}\right)^\ell&\approx&\left(\frac{\ell-ik_{||}r}{kr}\right)^\ell \nonumber \\
&=&\left(\frac{k_{\perp}-ik_{||}}{k}+O((k_{\perp}/k) (1/N))\right)^\ell.
\end{eqnarray}
The correction term above is small since $k_{\perp}/k<1$ and $N>>1$, so we have
\begin{eqnarray}
\left(\frac{\ell-i\sqrt{x^2-\ell^2}}{x}\right)^\ell&\approx&\left(\frac{k_{\perp}-ik_{||}}{k}\right)^\ell.
\end{eqnarray}
In the $k_{\perp}-k_{||}$ plane we may now make the following redefinition $k_{\perp}+ik_{||}=ike^{-i\varphi_k}$. This makes the above expression simply $e^{i\varphi_k \ell-\frac{\pi\ell}{2}}$. Thus, the final expression for the Bessel function becomes
\begin{eqnarray}
j_{\ell}(kr)&=&\frac{\sin{(k_{||}r+\varphi_{k}\ell-\pi\ell/2)}}{kr}.
\end{eqnarray}
Note that in the limit $kr>>\ell \,(k_{||}>>k_{\perp})$, $\varphi_k\approx 0$ and $k\approx k_{||}$, which recovers the well known approximation for the Bessel function in this regime.}

\section{Foregrounds}\label{A:fore}
{Foreground contamination in the flat-sky case is rather straightforward, owing to the fact that the foreground power spectrum has a simple expression (L17) which can be derived straightforwardly for a $T(\vec{r})=T(\hatr)$ given so,
\begin{eqnarray}
C_\ell^{\text{fg}}(k_{||})&=&C^{\text{fg}}_\ell \sinc^2{(k_{||}\Delta R/2)}\, ,
\end{eqnarray}
for a survey at mean comoving radius $R$ and survey width $\Delta R$, particularly for a tophat window function (which is the case we are concerning ourselves with in the SFB formalism). In particular, this expression is independent of $k_{\perp}$ since $C^{\text{fg}}_\ell$ is actually constant, i.e., one only needs to impose a cutoff $k_{||,\text{min}}$. Below, we will derive an expression to determine the appropriate methods for foreground cutoffs for SFB, which has a nontrivial expression for foreground power. For instance, it was shown in L17 that the foreground cuts for SFB will be scale-dependent (which is not the case for Plane-Parallel). Here, we will show that the correct cuts in SFB coordinates are given in terms of $k_{||,\rm min}$ by $k^2_{\rm min}=k_{||,\rm min}^2+\ell^2/r_{\rm max}^2$. Note that any radial integrals will be considered to be from $r_{\rm min}$ to $r_{\rm max}$, but we will not explicitly write the limits.}

{To start, consider the foreground signal in SFB (L17),
\begin{eqnarray}
\tau_{\ell n}C_{\ell n}^{\text{fg}} &=&  C_{\ell}^{\text{fg}}\frac{\left[\int dr\,r^2 j_{\ell}(k_{\ell n}r)\right]^2}{\int dr\, r^2 j_\ell^2(k_{\ell n}r)}\, ,
\end{eqnarray}
In particular, $\tau_{\ell n}C_{\ell n}^{\text{fg}}$ is some nontrivial function which is expected to produce a nontrivial $k_{\ell n, \text{min}}$ corresponding to the plane-parallel $k_{||,\text{min}}$}

{In particular we wish to examine $\tau_{\ell n}C_{\ell n}^{\text{fg}}$ in the regime $\ell>>1$ and $k_{\ell n}r\gtrsim\ell$ in order to obtain the correct $k_{\text{min}}$. Using the approximation for $j_{\ell}(kr)$ obtained in Appendix~\ref{A:saddle} we have
\begin{widetext}
\begin{eqnarray}
k_{\ell n}\int dr\,r^2 j_\ell(k_{\ell n}r) &=& \frac{\sin{(k_{||,\ell n}r_{\rm max}+\zeta)}-\sin{(k_{||,\ell n}r_{\rm min}+\zeta)}}{k_{||,\ell n}^2}+\frac{r_{\rm min}\cos{(k_{||,\ell n}r_{\rm min}+\zeta)}-r_{\rm max}\cos{(k_{||,\ell n}r_{\rm max}+\zeta)}}{k_{||,\ell n}} \nonumber \\
&=& R\Delta R\sinc{(k_{||,\ell n}\Delta R/2)}\sin{(k_{||,\ell n}R+\zeta)}+\frac{\Delta R}{k_{||,\ell n}}\sinc{(k_{||,\ell n}\Delta R/2)}\cos{(k_{||,\ell n}R+\zeta)}\nonumber \\
&&-\frac{\Delta R}{k_{||,\ell n}}\cos{(k_{||,\ell n}\Delta R/2)}\cos{(k_{||,\ell n}R+\zeta)}\, ,
\end{eqnarray}
\end{widetext}
where $k_{||,\ell n}^2\equiv k_{\ell n}^2-\ell^2/r_{\rm max}^2$ and $\zeta=\varphi_{k}\ell-\pi\ell/2$. Now since we have $k_{\ell n}r\approx \ell$, we have $k_{||,\ell n}\approx 0$ (to order 1/N). Thus, $\varphi_{k}\approx \pi/2$ so $\zeta\approx 0$. So we have
\begin{eqnarray}
k_{\ell n}\int dr\,r^2 j_\ell(k_{\ell n}r) &=& R\Delta R\sinc{(k_{||}\Delta R/2)}\sin{(k_{||,\ell n}R)}\nonumber \\
&&-R\Delta R \frac{\cos(k_{||}R)}{k_{||,\ell n}R}\nonumber \\
&&\times(\cos(k_{||}\Delta R/2)-\sinc(k_{||}\Delta R/2)).
\end{eqnarray}
However, note that in the limit $k_{||}\rightarrow 0$, particularly for the case $R\gg\Delta R$, the second term is subdominant. So to leading order, we may say
\begin{eqnarray}
k_{\ell n}\int dr\,r^2 j_\ell(k_{\ell n}r) &\approx& R\Delta R\sinc{(k_{||,\ell n}\Delta R/2)}\nonumber \\
&&\times \sin{(k_{||,\ell n}R)}.
\end{eqnarray}}

\begin{figure}
\begin{center}
\includegraphics[width=0.5\textwidth]{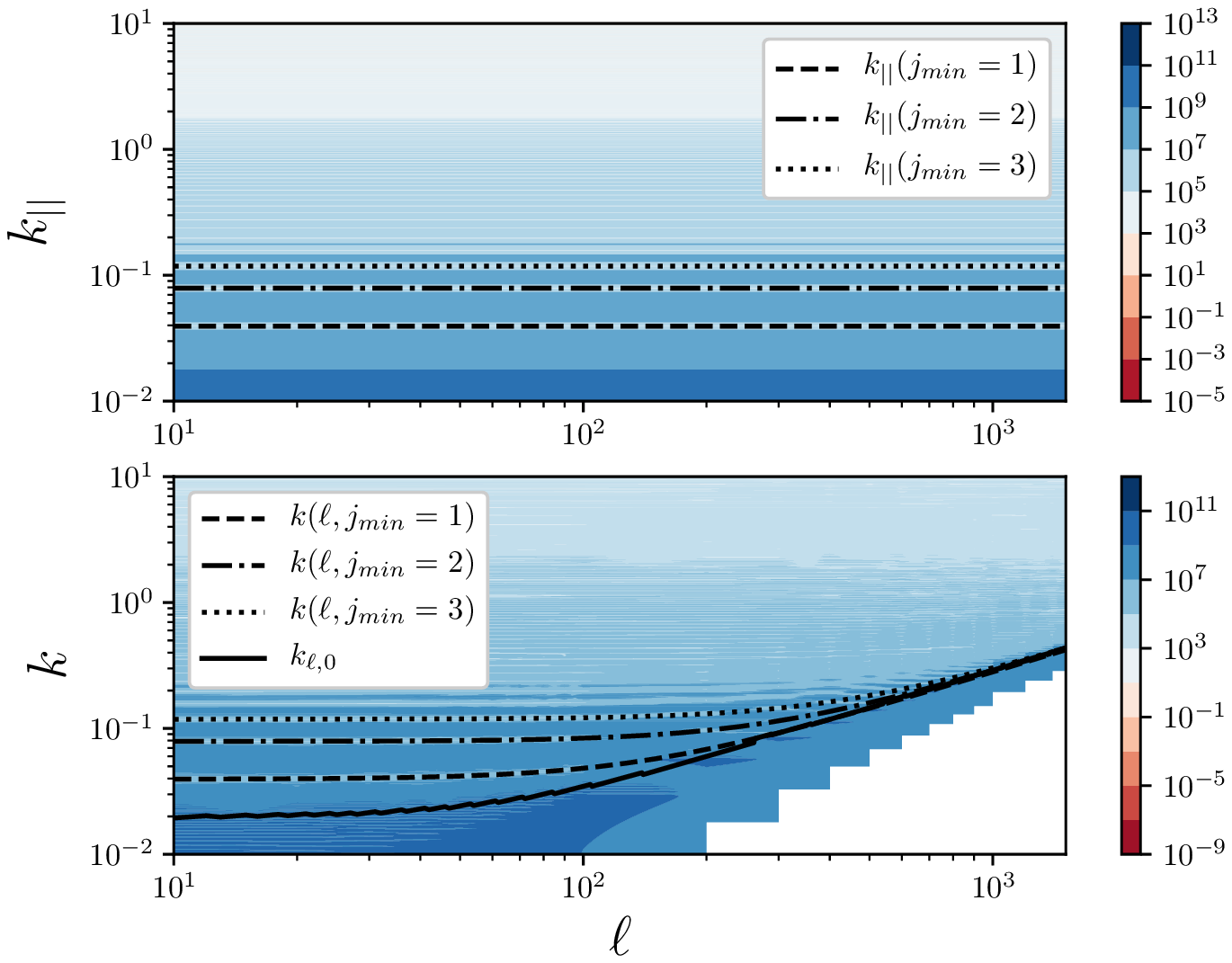}
\caption{\label{F:fore}  {$C_{\ell}^{\text{fg}}(k_{||})$ (above) and $C_{\ell n}^{\text{fg}}$ (below). The lines in in the PP plot show the contours of constant $C_{\ell}^{\text{fg}}(k_{||_j})$ for different choices of $j_{\text{min}}$. The contours in the plot for $C_{\ell n}^{\text{fg}}$ show the curve $k_{\text{min}}^2=k_{||,\text{min}}^2+\frac{\ell^2}{r_{\rm max}^2}$, with survey data fixed by $z=2\,,dz=0.16$. Note that $C_{\ell}^{\rm fg}=1$ in these plots.}}
\end{center}
\end{figure}

{Next we need the denominator which is just $W_{\ell \ell}(k,k)$, i.e., 
\begin{eqnarray}
\frac{2k^2}{\Delta R}W_{\ell \ell}(k,k) &\approx& \cos(0)\sinc(0)-\cos(2k_{||}R)\sinc(k_{||}\Delta R) \nonumber \\
&\approx& 1\, ,
\end{eqnarray}
since $k_{||}\approx 0$.  To complete, we have
\begin{eqnarray}
\tau_{\ell n}C_{\ell n}^{\text{fg}} &=& 2 C_{\ell}R^2\Delta R\sinc^2(k_{||,\ell n}\Delta R/2)\sin^2{(k_{||,\ell n}R)} \, ,
\end{eqnarray}
However, we may average out the oscillatory part of this expression as in L17, which gives exactly that $\tau_{\ell n}C_{\ell n}\approx R^2\Delta RC_{\ell}(k_{||,\ell n})$ when $k_{\ell n}R\gtrsim\ell$. This analysis indicates that the correct contours should be $k_{\text{min}}^2=k_{||,\text{min}}^2+\frac{\ell^2}{r_{\rm max}^2}$, as shown in Fig.~\ref{F:fore}.}

%%%%%%%%%%%%%%%%%%%%%%%%%%%%%%%%%%%%%%%%%%%%%%%%%%

% Don't change these lines
\bsp	% typesetting comment
\label{lastpage}
\end{document}